\PassOptionsToPackage{prologue,dvipsnames}{xcolor}
\documentclass[UKenglish]{article}

\usepackage[T1]{fontenc}
\usepackage[UKenglish]{babel}
\usepackage{graphicx}
\usepackage{amsmath}
\usepackage{amssymb}
\usepackage{relsize}
\usepackage{tikz-qtree}
\usepackage{tikz}
\usetikzlibrary{trees,arrows}
\usepackage{verbatim}
\usepackage{braket}
\usepackage{xargs}
\usepackage{xcolor}
\usepackage{xifthen} 
\usepackage{tikz}
\usepackage{etoolbox}
\usepackage[title]{appendix}
\usepackage{afterpage}
\usepackage{flafter}
\usepackage{float}
\usetikzlibrary{calc, shapes, backgrounds}
\usetikzlibrary{shapes.geometric, arrows}
\usetikzlibrary{calc, shapes, backgrounds}
\usetikzlibrary{shapes.geometric, arrows}
\usetikzlibrary{automata, arrows.meta, positioning}
\usepackage{tabu}
\usepackage[hidelinks]{hyperref}
\usepackage[utf8]{inputenc}
\usepackage[acronym]{glossaries}
\usepackage{csquotes}
\usepackage{orcidlink}
\usepackage{lineno}
\usepackage{soul}
\usepackage{wrapfig}
\usepackage{adjustbox}
\usepackage{caption}

\usepackage[percent]{overpic}

\newcolumntype{L}{>{\centering\arraybackslash}m{5cm}}
\newcolumntype{Z}{>{\centering\arraybackslash}m{1.6cm}}

\usepackage[textwidth=17cm,textheight=21.7cm]{geometry}
\usepackage{tikz}
\usetikzlibrary{quantikz2}
\usepackage[shortlabels]{enumitem}
\usepackage{diagbox}
\usepackage{graphicx}
\usepackage{subcaption}
\usepackage{booktabs}
\usepackage[backend=biber,style=nature]{biblatex}
\usepackage{orcidlink}

\newsavebox\curwrapfig
\makeatletter
\long\def\wrapfiguresafe#1#2#3{%
  \sbox\curwrapfig{#3}%
  \par\penalty-100%
  \begingroup 
    \dimen@\pagegoal \advance\dimen@-\pagetotal 
    \advance\dimen@-\baselineskip 
    \ifdim \ht\curwrapfig>\dimen@ 
      \break%
    \fi%
  \endgroup%
  \begin{wrapfigure}{#1}{#2}%
    \usebox\curwrapfig%
  \end{wrapfigure}%
}
\makeatother

\title{Quantum networking with advances in fiber technology}
\vspace{-6mm}
\author{Prateek Mantri$^{1*}\orcidlink{0009-0005-3223-9731}$, Michael S. Bullock$^{1,2}$, Aditya Tripathi$^{2}$, Robert Kwolek$^{2}$, \\Rajveer Nehra$^{1,2,3}$, and Don Towsley$^{1}$}
\date{\small \textit{$^{1}$Robert and Donna Manning College of Information and Computer Science, \\ \small University of Massachusetts Amherst, 140 Governor's Drive, Amherst, MA 01003, USA\\ \small
$^{2}$Department of Electrical and Computer Engineering, University of Massachusetts, Amherst, MA 01003, USA\\ \small
$^{3}$ Department of Physics, University of Massachusetts Amherst, MA 01003, USA
\\\small $^*$Email: pmantri@cs.umass.edu}}

\vspace{-12pt}

\begin{document}
\maketitle
\begin{abstract}
Recent comparisons of quantum repeater protocols have highlighted the strong near-term potential of multiplexed two-way architectures for long-distance quantum communication~\cite{mantri2025comparing}. At the same time, advances in hollow-core fiber (HCF) technology~\cite{NumkamFokoua23hcfreview,Petrovich2025} motivate a re-examination of the physical transmission medium as an architectural lever in quantum network design. In this work, we compare emerging anti-resonant HCFs against conventional silica single-mode fibers (SMFs) in multiplexed two-way quantum repeater networks. We evaluate their performance under both telecom and memory-native transmission, accounting for frequency-conversion overheads, coupling efficiencies, memory decoherence, and operational noise. 

We find that HCF significantly outperforms SMF across a wide range of regimes. With memory-native transmission, HCF yields up to an order of magnitude improvement in secret-key rate per channel use under realistic conversion efficiencies. Even at telecom wavelengths, HCF enables larger optimal repeater spacing, improving rate--cost tradeoffs and reducing repeater requirements. We further quantify the role of memory quality, hardware efficiency, detector and conversion losses, and two-qubit gate noise in shaping these gains. These results show that recent advances in HCF materially expand the design space of practical terrestrial quantum repeater networks.
\end{abstract}


\section{Introduction}

Quantum information processing (QIP) offers fundamentally new capabilities beyond classical information systems, including provably secure communication, distributed quantum sensing, and scalable modular quantum computing~\cite{pirandola2020advances, pan_evolution_2023, degen_quantum_2017, caleffi_distributed_2024}. A unifying feature across these applications is the need to distribute quantum states---most notably entanglement---across distant nodes. In contrast to classical networks, where information can be copied, amplified, and routed arbitrarily, quantum information must be transmitted while preserving fragile quantum correlations. As a result, networking is not merely a supporting infrastructure for QIP, but a central component that enables these capabilities at scale. Realizing large-scale quantum networks therefore requires reliable methods for generating, distributing, and maintaining entanglement across long distances~\cite{wehner18quantuminternet, Kimble_2008}.

Loss is one of the most fundamental challenges in quantum communication~\cite{guha2015rate, takeoka_capacity_2014}.  The probability that a photonic quantum state survives transmission decreases rapidly with distance, making loss the dominant factor limiting the scale, rate, and reach of quantum networks. Furthermore, unlike classical signals, quantum information cannot be amplified in transit, owing to the no-cloning theorem. In fiber-based links, photon transmission follows an exponential decay with distance,
\begin{align}
    p_\mathrm{succ} \propto \exp\!\left(-\frac{L_0}{L_\mathrm{att}}\right),
\end{align}
where $L_0$ is the link length and $L_\mathrm{att}$ is the attenuation length of the optical channel. Even in modern silica single-mode fibers (SMFs), which achieve attenuation coefficients near $0.2\,\mathrm{dB/km}$ around $1550\,\mathrm{nm}$, corresponding to attenuation lengths of roughly $20$--$22\,\mathrm{km}$, the probability of successful transmission still falls exponentially with distance, making direct long-distance quantum communication impractical.  This limitation is formalized by repeaterless rate--loss bounds such as the Takeoka--Guha--Wilde (TGW) bound and the Pirandola--Laurenza--Ottaviani--Banchi (PLOB)bound~\cite{takeoka_capacity_2014,PLOBBound,tgwBound}.

A substantial body of work has therefore focused on mitigating this loss bottleneck. Proposed approaches include quantum repeaters based on entanglement generation, entanglement swapping, and distillation; encoded and error-corrected transmission schemes; improved memories, emitters, and detectors; and multiplexing across time, frequency, and space~\cite{azuma2023quantum,muralidharan2016optimal,MAT2015Qrepeaters,dur_entanglement_2007,briegelQuantumRepeatersCommunication1998,briegelQuantumRepeatersRole1998}. Within repeater-based networking, two broad architectural paradigms have emerged. Two-way repeater protocols generate elementary entanglement probabilistically and rely on heralding before performing higher-level operations, making them relatively robust to loss. In contrast, one-way (third-generation) repeater architectures use quantum error correction to avoid heralding delays but require substantially more demanding hardware~\cite{muralidharan2016optimal,muralidharan2014ultrafast}. As a result, two-way architectures remain especially attractive for near-term implementations~\cite{mantri2025comparing}.

Alternative approaches seek to circumvent terrestrial fiber losses altogether using free-space or satellite-based quantum links~\cite{yin2017satellite,liao2017satellite,ren2017ground,li2022space}. Related proposals such as vacuum beam guides also aim to reduce propagation loss while avoiding some of the constraints of conventional fiber~\cite{huang2024vacuum}. While these approaches are promising, they introduce additional challenges including intermittency, atmospheric effects, pointing and tracking requirements, or significant infrastructure cost~\cite{jha2025towards,huang2024vacuum}. By contrast, terrestrial optical fiber remains inexpensive, continuously available, mature, and deeply integrated into existing communication infrastructure. It therefore remains one of the most practical candidates for large-scale quantum communication, provided the physical layer can support sufficiently low loss and high performance.

Recent advances in hollow-core fiber (HCF) technologies show promise in mitigating some of the constraints listed above~\cite{Petrovich2025,Sakr:21,Adamu:24,NumkamFokoua23hcfreview}. The goal of this paper is to compare conventional silica SMF with HCF-based schemes in large-scale multiplexed two-way repeater architectures. The central question is not merely whether HCF offers lower propagation loss than SMF in isolation, but whether it changes the preferred operating regime of an end-to-end quantum network once wavelength choice, coupling, frequency conversion, detection, and repeater spacing are all accounted for. In particular, we revisit the common assumption that terrestrial quantum repeater networks should operate near the telecom wavelength of 1550~nm. While this choice is natural for conventional silica fiber, many quantum memories and emitters operate at shorter visible or near-infrared wavelengths, so telecom operation often requires two-stage aggregate frequency conversion and its associated insertion loss and hardware complexity~\cite{ma2017optical, hannegan2022entanglement, knaut2024entanglement,carosini2026quantum}. In addition to attenuation, physical-layer effects such as mismatch between memory/emitter wavelengths and optimal fiber-transmission wavelengths, phase noise, chromatic dispersion, and spontaneous Raman scattering further constrain the temporal and spectral packing of photonic wavepackets, thereby limiting achievable multiplexing and end-to-end rates. This wavelength question is becoming increasingly important as quantum-network hardware matures. Progress in quantum-dot single-photon sources~\cite{senellart2017high,carosini2026quantum}, low-loss electro-optic programmable interferometers for ultrafast switching~\cite{sund2023high,dong2023synchronous}, and efficient near-infrared single-photon detection~\cite{an2025,excelitas} points toward architectures in which sources, switches, detectors, memories, and transmission media must be designed as a coupled system. Recent work has already demonstrated this direction by engineering HCF for low-loss transmission at quantum-dot wavelengths while maintaining compatibility with telecom-band classical channels~\cite{carosini2026quantum}. Together with broader advances in HCF, these developments motivate treating the fiber platform and operating wavelength as co-optimized architectural degrees of freedom, rather than simply inheriting the 1550~nm operating point of classical optical communication.

In this work, we evaluate SMF- and HCF-based repeater networks under a common end-to-end modeling framework, explicitly incorporating physical-layer transmission, interface losses, and system-level architectural choices. Our goal is to determine whether recent advances in HCF materially shift the optimal design of practical terrestrial quantum repeater networks. 

The remainder of the paper is organized as follows. Section~\ref{sec:background} provides the relevant physical-layer and system-level background. Section~\ref{sec:system_description} describes the system under study. Section~\ref{sec:Methods} presents the modeling framework and simulation methodology. Section~\ref{sec:Results} then uses these tools to evaluate the network-level implications of recent HCF proposals. Section~\ref{sec:Conclusions} concludes the paper and outlines directions for future work.


\section{Background}
\label{sec:background}

\subsection{End-to-end loss in quantum networks}

Propagation loss in the transmission channel is only one component of the total loss budget in a quantum network. In practice, photons encounter multiple lossy interfaces throughout the entanglement distribution process: emission from quantum memories, coupling into guided optical modes, frequency conversion, propagation through the channel, and final detection efficiency. The success probability of an elementary entanglement-generation attempt is therefore governed by the product of efficiencies across all of these processes~\cite{guha2015rate,muralidharan2016optimal}.

This broader systems perspective is essential when evaluating repeater architectures. Improvements in one component of the photonic interface can easily be offset by losses elsewhere, so a transmission medium should not be assessed solely by its propagation coefficient. Instead, the relevant question is how the medium interacts with wavelength choice, interface efficiency, detector compatibility, and repeater architecture to determine end-to-end network performance.

\subsection{Wavelength mismatch and hardware constraints}

Many leading quantum memory platforms operate at memory-native wavelengths outside the telecom band, often in the visible or near-infrared~\cite{ma2017optical, hannegan2022entanglement, knaut2024entanglement, carosini2026quantum}. Bridging this mismatch typically requires quantum frequency conversion, which introduces additional loss and system complexity~\cite{zaske2012visible,tamura2018two,cohen2026compact,li2022frequency}. Recent remote-entanglement demonstrations using solid-state and trapped-ion quantum-memory nodes have likewise relied on frequency conversion to interface non-telecom optical transitions with low-loss telecom-fiber channels~\cite{stolk2024metropolitan,liu2026long}. Consequently, the conventional assumption that quantum networks should always operate at telecom wavelengths is not automatically optimal once end-to-end losses are taken into account.

Detection technology further sharpens this tradeoff. Superconducting nanowire single-photon detectors (SNSPDs) offer high efficiency at telecom wavelengths but require cryogenic operation~\cite{reddy2020,reddy2022broadband}, whereas room-temperature avalanche photodiodes (APDs) typically perform best at shorter wavelengths~\cite{an2025,excelitas}. Consequently, transmitting closer to a memory-native wavelength may sometimes reduce overall system penalty even when the propagation medium itself is less favorable at that wavelength.

\subsection{Limitations of conventional silica fibers}

Conventional silica SMFs impose important constraints beyond attenuation alone. Because they guide light through total internal reflection in solid glass, their transmission characteristics are fundamentally shaped by material-induced effects. In particular, the attenuation floor in silica fibers is dominated by Rayleigh scattering arising from intrinsic density fluctuations in the glass. This contribution scales approximately as $\lambda^{-4}$, leading to substantially higher loss away from the telecom window~\cite{agrawal2012nonlinear}. As a result, SMFs perform very well near $1550\,\mathrm{nm}$ but become increasingly unfavorable at shorter, memory-native wavelengths.

Silica fibers are also limited by additional material-driven impairments. Chromatic dispersion causes wavelength-dependent pulse broadening, which complicates broadband transmission and highly multiplexed operation. Spontaneous Raman scattering from co-propagating classical channels generates broadband noise that can contaminate weak quantum signals and hinder classical-quantum coexistence~\cite{thomas2023designing}. These effects are not merely secondary implementation details: they directly constrain usable wavelength ranges, multiplexing strategies, coexistence models, and ultimately the architectural design of practical quantum networks.

\subsection{Hollow-core fiber as an alternative physical layer}

\begin{figure}[!htbp]
  \centering
  \begin{subfigure}[t]{0.30\textwidth}
    \centering
    \includegraphics[width=\linewidth]{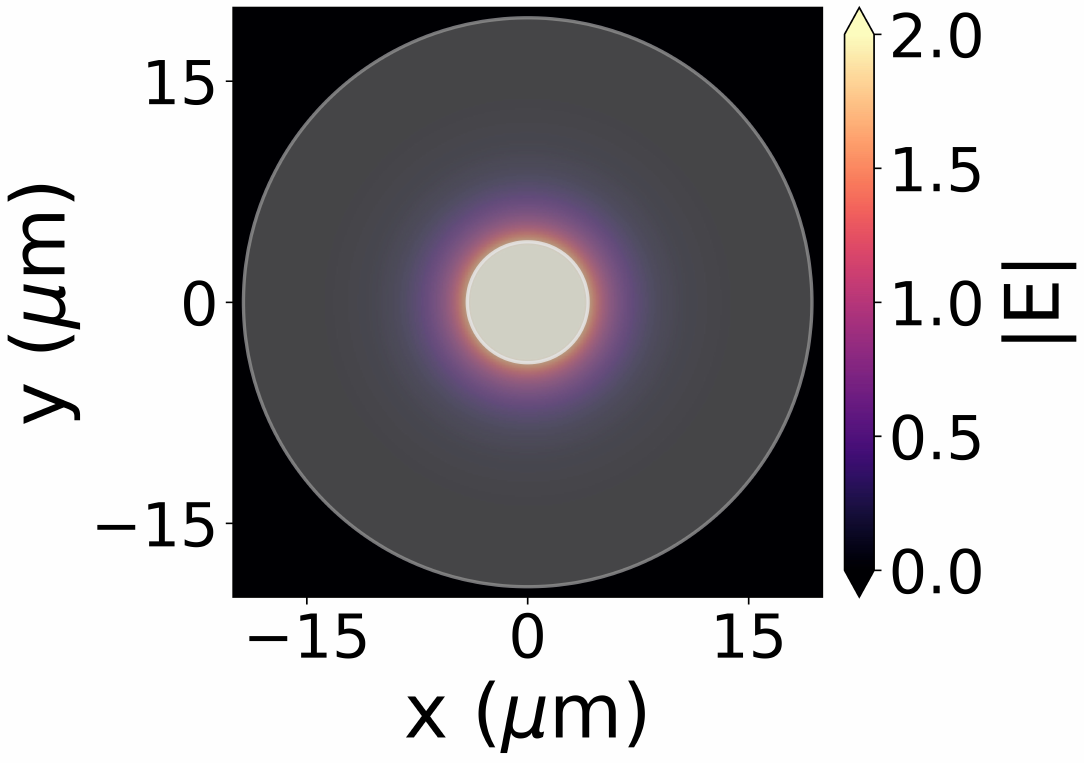}
    \caption{}
    \label{fig:smf}
  \end{subfigure}\hfill
  \begin{subfigure}[t]{0.30\textwidth}
    \centering
    \includegraphics[width=\linewidth]{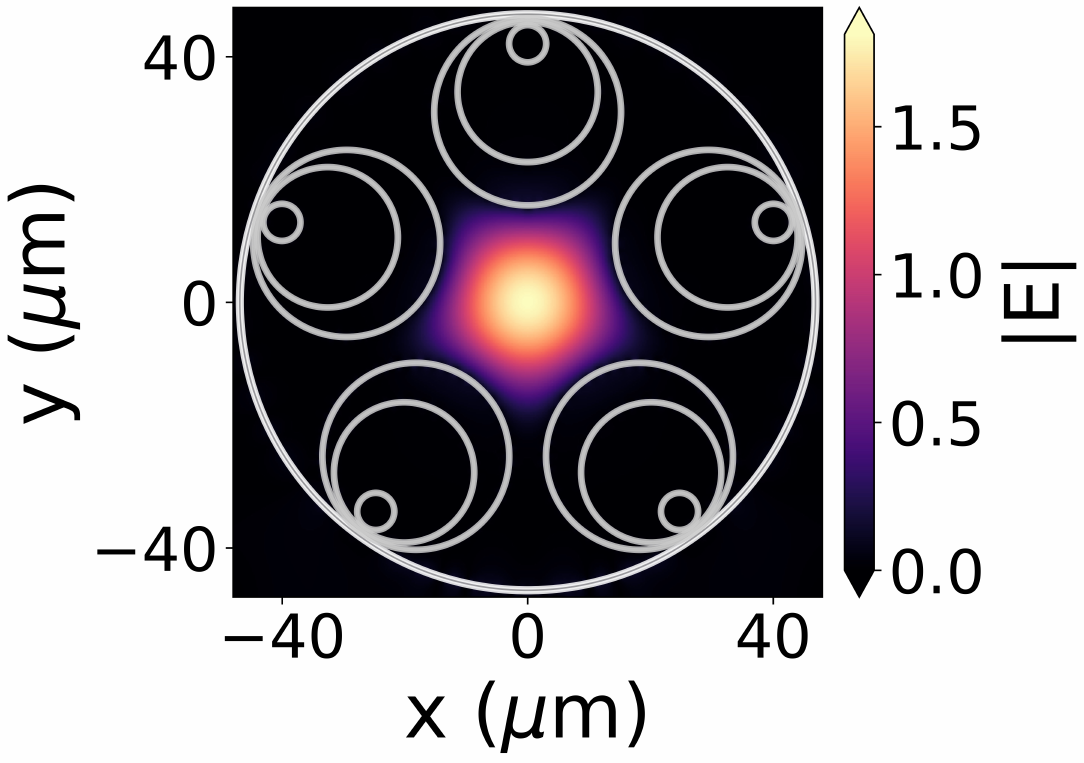}
    \caption{}
    \label{fig:hcf}
  \end{subfigure}\hfill
  \begin{subfigure}[t]{0.39\textwidth}
    \centering
    \begin{overpic}[width=\linewidth]{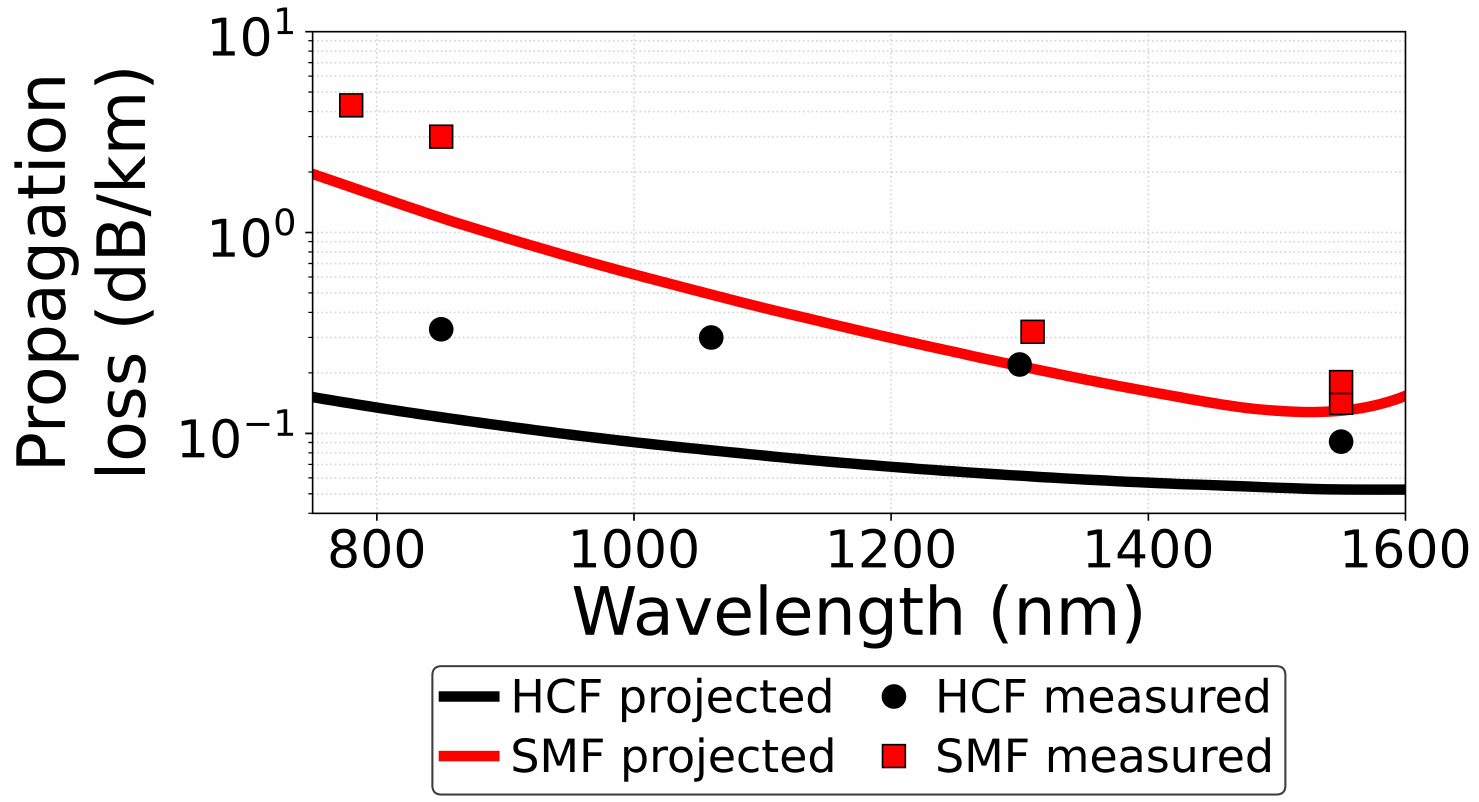}
      \put(86,23.3){\tiny\color{black}\cite{Petrovich2025}}
      \put(46.5,28.5){\tiny\color{black}\cite{Sakr:21}}
      \put(66.5,26.5){\tiny\color{black}\cite{Sakr:21}}
      \put(32,32){\tiny\color{black}\cite{Adamu:24}}
      \put(84,27.5){\tiny\color{red}\cite{Sato24smf}}
      \put(88,30.4){\tiny\color{red}\cite{corning_smf28_productinfo}}
      \put(72,32){\tiny\color{red}\cite{corning_smf28_productinfo}}
      \put(26,47){\tiny\color{red}\cite{corning_hi780_spec}}
      \put(32,45){\tiny\color{red}\cite{corning_hi780_spec}}
    \end{overpic}
    \caption{}
    \label{fig:loss}
  \end{subfigure}
  \caption{\textbf{Physical-layer comparison of conventional silica single-mode fiber (SMF) and hollow-core fiber (HCF).} (a) Cross section and fundamental guided mode of a step-index SMF. (b) Cross section and fundamental guided mode of a double-nested anti-resonant hollow-core fiber. (c) Propagation loss across relevant wavelengths for representative SMF and HCF technologies, using measured data from prior work~\cite{Petrovich2025, Sakr:21, Adamu:24, Sato24smf, corning_hi780_spec, corning_smf28_productinfo} and projected achievable performance from~\cite[Fig.~40]{NumkamFokoua23hcfreview}. Attenuation values used in the simulations are $0.150$~dB/km (SMF, 1550~nm), $0.055$~dB/km (HCF, 1550~nm), and $0.180$~dB/km (HCF, 780~nm).}
  \label{fig:smf_hcf_loss}
\end{figure}
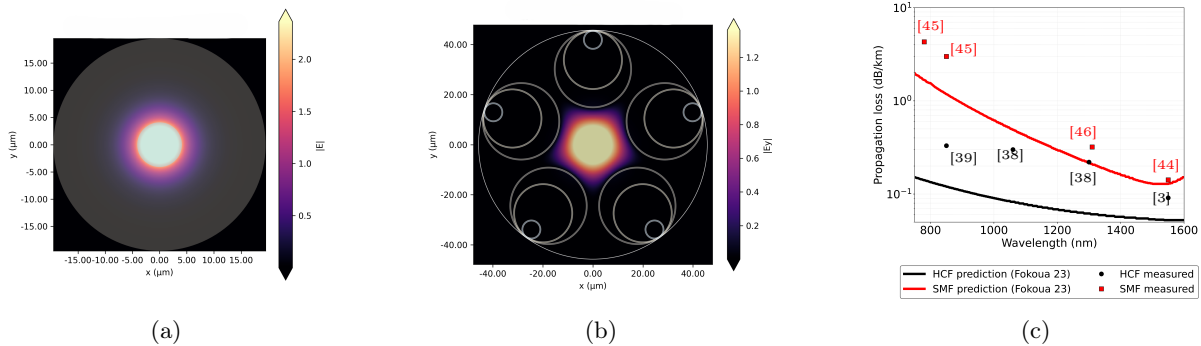

Hollow-core fibers guide light predominantly through air rather than solid glass, reducing the interaction between the optical mode and the material medium. As illustrated in Fig.~\ref{fig:smf_hcf_loss}, conventional SMFs and modern HCFs differ not only in cross-sectional structure and mode confinement, but also in their wavelength-dependent loss profiles. This leads to qualitatively different loss, dispersion, and noise characteristics compared with conventional SMFs, while also suppressing several of the mechanisms that limit solid-core silica transmission, including material-induced nonlinearities and Raman noise~\cite{Petrovich2025,Sakr:21,Adamu:24,NumkamFokoua23hcfreview}.

Early photonic bandgap HCFs demonstrated the possibility of non-total-internal-reflection guidance, but suffered from relatively narrow bandwidths and higher scattering losses. Modern anti-resonant designs overcome many of these limitations by using thin-walled glass structures that reflect non-resonant wavelengths back into the hollow core, enabling broadband and low-loss guidance~\cite{russell2003photonic,Benabid02HCF,Poletti14nanf,NumkamFokoua23hcfreview}. In particular, nested and double-nested anti-resonant nodeless fibers (DNANFs) reduce scattering and confinement leakage, and have achieved record-low propagation losses together with improved modal confinement~\cite{Poletti14nanf,Petrovich2025}.

These developments are especially significant for quantum networking. Because light in HCF propagates primarily in air, Raman scattering is dramatically suppressed relative to solid silica, opening the possibility of cleaner classical-quantum coexistence and lower-noise transmission~\cite{clark25coexhcf}. HCFs also offer low dispersion and a broader usable wavelength range than conventional SMFs, making them attractive not only at telecom wavelengths but also closer to the native emission wavelengths of many quantum memories and matter-based quantum computing and sensing platforms. Recent DNANF implementations have reported propagation losses below $0.1\,\mathrm{dB/km}$ near $1550\,\mathrm{nm}$ in laboratory settings~\cite{Petrovich2025}, surpassing the loss levels historically associated with conventional solid-core silica transmission.

More importantly for network design, HCF may alter the entire architectural trade space by changing the balance among propagation loss, wavelength choice, conversion overhead, detector compatibility, and repeater spacing. In that sense, HCF is not merely a lower-loss drop-in replacement for SMF; it creates the possibility of a different optimum for end-to-end quantum networking systems.

\section{System Description}\label{sec:system_description}

\begin{figure}[htb!]
    \centering
    \includegraphics[width=0.95\linewidth, trim = 0cm 0cm 0cm 0cm , clip]{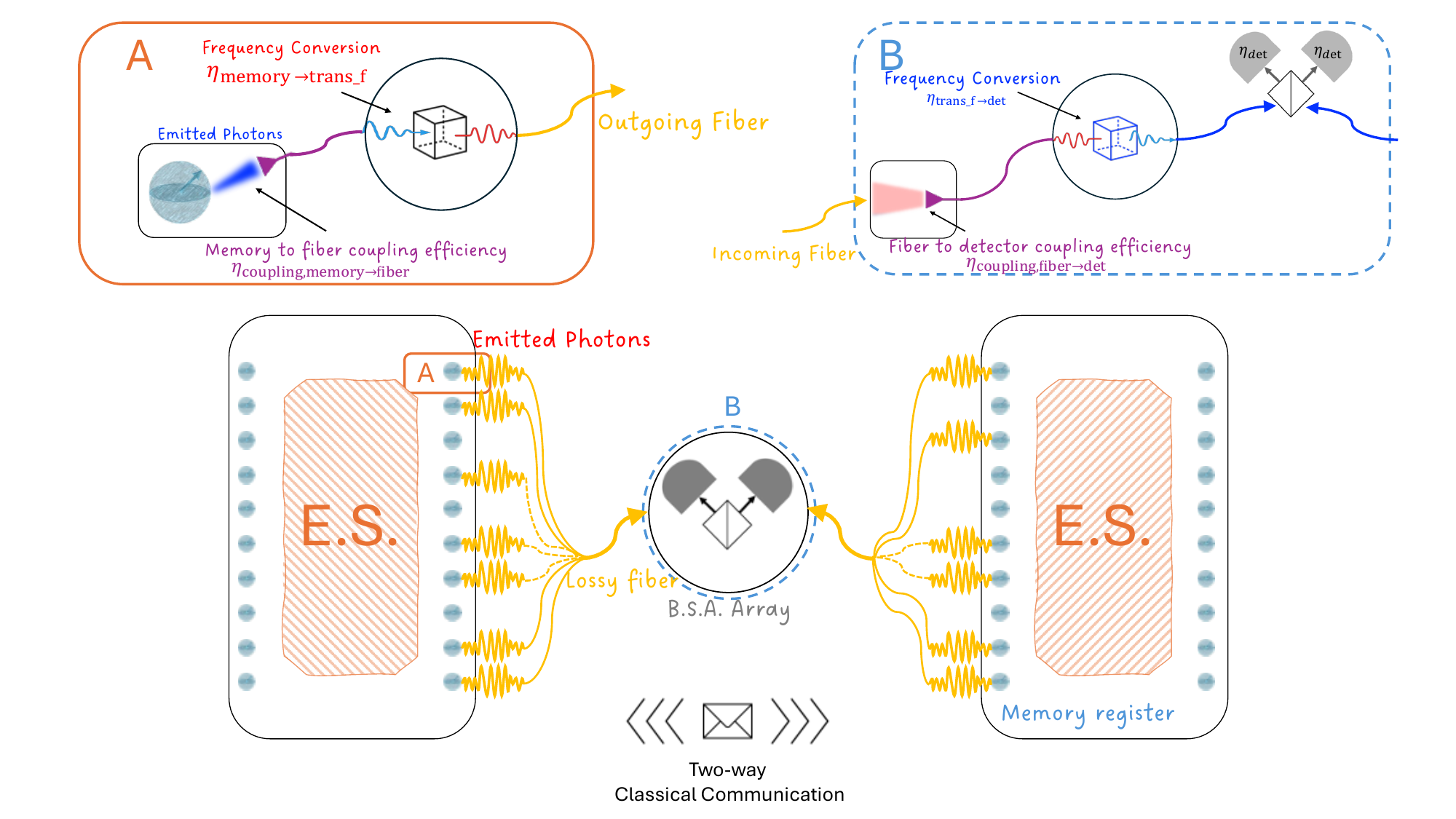}
    \caption{
    \textbf{Architecture for multiplexed entanglement generation and entanglement swapping between neighboring repeater nodes.} Each node contains a register of quantum memories coupled to photonic emitters that generate photons entangled with the stored memory qubits. 
    (A) Photons emitted from the memory register are coupled into the outgoing optical fiber with efficiency $\eta_{\mathrm{coupling,\,memory\rightarrow fiber}}$ and may undergo frequency conversion with efficiency $\eta_{\mathrm{memory\rightarrow trans\_f}}$ before transmission through the channel. 
    (B) Photons arriving from the channel are coupled from the fiber into the receiver optics with efficiency $\eta_{\mathrm{coupling,\,fiber\rightarrow det}}$ and may undergo frequency conversion with efficiency $\eta_{\mathrm{trans\_f\rightarrow det}}$ before detection with detector efficiency $\eta_{\mathrm{det}}$. 
    Photons from neighboring nodes interfere at the midpoint Bell-state analyzer (BSA) array. Successful Bell-state measurements herald the creation of entanglement between remote memory qubits stored in the two nodes. Classical signals from the BSA are communicated back to the nodes via a two-way classical channel to identify successful entanglement events. Once entanglement is established, repeaters perform the entanglement swapping (E.S.) operations to propagate the link.
}
\label{fig:system_desc_repeater}
\end{figure}

\begin{figure}[htb!]
    \centering
    \includegraphics[width=\linewidth, trim = 0cm 0cm 0.25cm 0cm, clip]{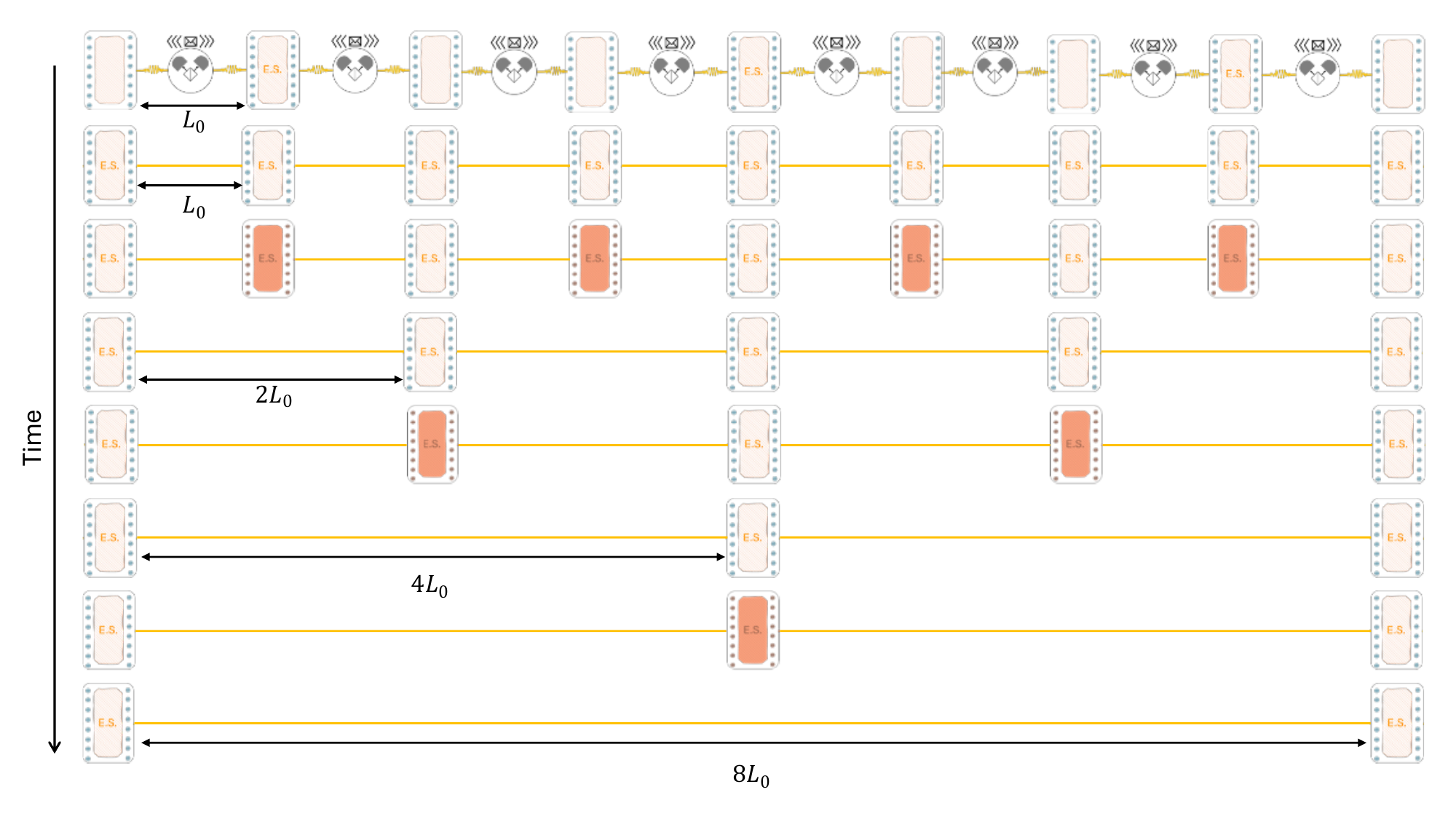}
    \caption{
    \textbf{Time evolution of the nested entanglement swapping protocol used to extend entanglement across a chain of repeater nodes.} Elementary links of length $L_0$ are first generated between neighboring nodes. Once two adjacent elementary links are available, an entanglement swapping (E.S.) operation is performed at the intermediate repeater, creating a longer link of length $2L_0$. The same procedure is recursively applied: swapping operations at higher levels combine neighboring links to produce entangled connections of length $4L_0$, $8L_0$, and ultimately longer end-to-end links. The vertical axis represents time, illustrating that entanglement generation attempts occur in parallel across elementary links, while swapping operations are triggered whenever the required shorter links become available. Highlighted repeater modules indicate nodes where entanglement swapping operations are performed to extend the entanglement length.
}
    \label{fig:system_desc_nested}
\end{figure}

We consider a linear quantum repeater chain composed of $N$ elementary links connecting Alice and Bob, with repeater stations placed between adjacent links. Each repeater is equipped with a large ensemble of optically active quantum memories (or emitters) capable of parallel entanglement generation through spatial or time-bin multiplexing. The overall modeling framework closely follows that developed in Mantri \textit{et al.} 2025~\cite{mantri2025comparing} and adopts the same physical-layer assumptions and noise models. Fig.~\ref{fig:system_desc_repeater} provides a schematic description of the elementary-link generation process, while Fig.~\ref{fig:system_desc_nested} illustrates the link propagation strategy employed in our system.

\subsection{Architectural Assumptions}

\paragraph{Unconstrained memory regime}
We operate in a regime where memory availability is not the primary bottleneck. Each elementary link supports $M$ parallel entanglement-generation attempts per burst, where a burst denotes one synchronized round of simultaneous elementary-link generation attempts across the available multiplexed channels~\cite{razavi_physical_2009, munro_quantum_2010, muralidharan2016optimal,chen_zero-added-loss_2023}. Each successful attempt produces one elementary Bell pair between neighboring nodes, and multiple simultaneous successes per link are retained (see Appendix~\ref{appendix:recursive_calcs} for details). Sufficient quantum memories are assumed available to store all successfully generated Bell pairs and intermediate distilled states. This reflects an architectural setting expected in large-scale deployments where memory scaling is feasible and classical latency associated with operations---not memory scarcity---dominates performance~\cite{mantri2025comparing}.

\paragraph{Elementary link generation}
Elementary links are generated using a two-photon, meet-in-the-middle remote-entanglement-generation protocol. Emitters at neighboring repeaters create photon--matter entangled states, and the emitted photons interfere at a midpoint Bell-state analyzer (BSA). Successful detection heralds entanglement between the remote memories. The success probability of a single elementary attempt is
\begin{equation}
    \pi_0(\lambda)
    =
    \frac{1}{2}\eta_c^2
    e^{-L_0/L_{\mathrm{att}}(\lambda)},
    \label{eq:p_succ_elem}
\end{equation}
where $L_0$ is the inter-repeater spacing, $L_{\mathrm{att}}(\lambda)$ is the fiber attenuation length, which depends on the transmission medium and wavelength, and $\eta_c$ is the effective photon coupling efficiency.

The parameter $\eta_c$ captures all non-fiber losses that occur between photon emission and successful detection at the midpoint station:
\begin{align*}
    \eta_c := &\;
    \eta_\mathrm{memory\text{-}emission}
    \cdot
    \eta_{\mathrm{coupling, memory\to fiber}}
    \cdot
    \eta_{\mathrm{memory_f}\to\mathrm{trans_f}}
    \cdot
    \eta_{\mathrm{trans_f}\to\mathrm{det_f}}
    \cdot
    \eta_{\mathrm{fiber\to det, coupling}}
    \cdot
    \eta_\mathrm{det}.
\end{align*}
These terms represent, respectively, the efficiency of photon emission from the memory, coupling into the transmission fiber, frequency conversion between the memory and transmission wavelengths, frequency conversion at the receiver, coupling from the fiber into the detection apparatus, and the quantum efficiency of the detector.

We group together device-dependent efficiencies that are largely independent of the transmission medium to define the aggregate hardware efficiency
\begin{align*}
    \eta_\mathrm{hardware}
    :=
    \eta_\mathrm{memory\text{-}emission}
    \cdot
    \eta_{\mathrm{fiber\to det, coupling}}
    \cdot
    \eta_\mathrm{det}.
\end{align*}
The remaining factors correspond to memory--fiber coupling and frequency conversion, giving
\begin{align*}
    \eta_c
    &=
        \eta_\mathrm{hardware}
        \;
        \eta_{\mathrm{coupling, memory\to fiber}}
        \;
        \eta_{\mathrm{memory_f}\to\mathrm{trans_f}}
        \;
        \eta_{\mathrm{trans_f}\to\mathrm{det_f}}
        \\
    &=
        \eta_\mathrm{hardware}
        \;
        \eta_{\mathrm{coupling, memory\to fiber}}
        \;
        \eta_\mathrm{conv},
\end{align*}
where
\begin{align*}
    \eta_\mathrm{conv}
    :=
    \eta_{\mathrm{memory_f}\to\mathrm{trans_f}}
    \eta_{\mathrm{trans_f}\to\mathrm{det_f}},
\end{align*}
is the aggregate efficiency of the two frequency-conversion stages.

The two-photon protocol considered here does not require stabilization of the absolute optical phase accumulated independently along the two remote paths, which is advantageous for the highly multiplexed architecture modeled in this work. Single-photon-interference schemes provide an important alternative and can exhibit more favorable transmission-loss scaling because successful heralding requires the detection of only one photon~\cite{sangouard2011quantum, yoshida2024multiplexed}. However, they generally impose more stringent requirements on long-baseline optical-phase stability than two-photon-interference schemes~\cite{zhao2007robust,yoshida2024multiplexed}. Their implementation also requires phase coherence and optical-mode indistinguishability to be maintained across the interfering modes~\cite{minavr2007phase, beukers2024remote}. When many temporal, spectral, or spatial modes are multiplexed, satisfying these requirements across all modes may introduce additional calibration, control, and stabilization overhead~\cite{yoshida2024multiplexed}. A quantitative comparison between one- and two-photon elementary-link protocols would therefore require different physical-layer models, together with distinct phase-stabilization, calibration, and control protocols. Such a comparison is beyond the scope of the present work and will be considered in a separate study.

\paragraph{Detection architecture}
For the baseline architecture considered throughout this work, we assume avalanche photodiode (APD) detection. Although superconducting nanowire single-photon detectors (SNSPDs) provide higher detection efficiencies at telecom wavelengths, they require cryogenic operation and additional infrastructure. APDs therefore provide a practical baseline due to their maturity, availability, and lower operational complexity. We separately evaluate the impact of direct telecom-band SNSPD detection in Sec.~\ref{sec:results_gatenoise} to assess how the choice of detector technology affects the relative performance of HCF and silica architectures.

\paragraph{Choice of transmission wavelength}

We assume a memory-native emission wavelength of $780\,\mathrm{nm}$, representative of near-$800\,\mathrm{nm}$ quantum-memory emission wavelengths. Elementary links can therefore be generated either through direct transmission at the memory wavelength or through telecom-band transmission at $1550\,\mathrm{nm}$ using two-stage aggregate quantum frequency conversion. 
For direct transmission at the memory wavelength we have
\begin{align*}
\pi_0{(780)} &=
\frac{1}{2}
\left(
    \eta_\mathrm{hardware}
    \;
    \eta_{\mathrm{coupling, memory\to fiber}}
\right)^2
\exp\!\left(-\frac{L_0}{L_{\mathrm{att}}(780)}\right),
\end{align*}
\noindent while telecom transmission includes frequency conversion losses,
\begin{align*}
\pi_0{(1550 )}
=
\frac{1}{2}
\left(
        \eta_\mathrm{hardware}
        \;
        \eta_{\mathrm{coupling, memory\to fiber}}
        \;
        \eta_\mathrm{conv}
\right)^2
\exp\!\left(-\frac{L_0}{L_{\mathrm{att}}(1550)}\right).
\end{align*}

\noindent For each elementary link we select the transmission wavelength that maximizes the link success probability,
\begin{align*}
    \pi_0 = \max\left(\pi_0{(780)},\,\pi_0{(1550)}\right).
\end{align*}

\begin{wrapfigure}{R}{0.52\textwidth}
  \centering
  \includegraphics[width=0.5\textwidth]{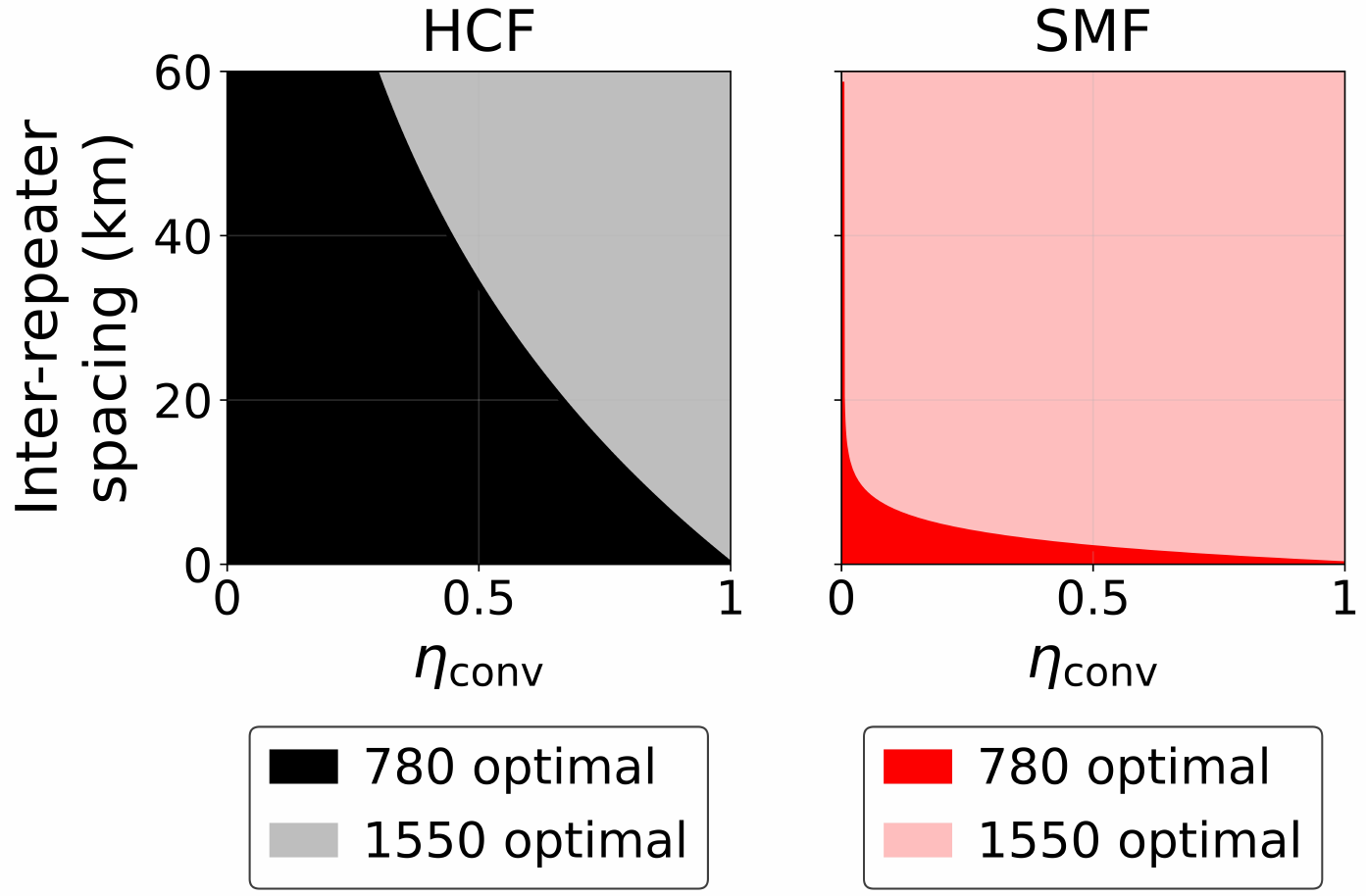}
	\caption{\textbf{HCF and SMF optimal strategy across frequency conversion efficiency and inter-repeater spacing parameter space.} Plots show the regions of the parameter space for which direct transmission at the quantum memory wavelength outperforms telecom-band transmission with two-stage frequency conversion. Darker regions for each plot indicate where direct transmission at 780 nm is optimal. }
  \label{fig:conv}
  \vspace{-0.5em} 
\end{wrapfigure}
\noindent This adaptive selection is evaluated independently for each elementary-link length $L_0$ in the simulation. While silica SMF can guide light at $780$--$800\,\mathrm{nm}$, attenuation at these wavelengths is prohibitively high, resulting in impractically short repeater spacings. This is reflected in Fig.~\ref{fig:conv}, where the memory-wavelength region is effectively excluded for SMF across the parameter range of interest. Accordingly, we do not consider SMF at the memory wavelength, and instead restrict attention to telecom-band transmission in SMF or direct transmission over low-loss media such as HCF. 

\paragraph{Nested entanglement swapping and recursive formulation.}
Long-distance entanglement is established via nested entanglement swapping across the repeater chain. Elementary Bell pairs generated between neighboring nodes (level $0$) are combined using Bell-state measurements at intermediate repeater nodes to form longer-distance entangled links. At level $t$, each successful swapping operation produces an entangled pair spanning $2^t$ elementary links, inducing a hierarchy in which longer links are recursively constructed from pairs of shorter links at the preceding level (Fig.~\ref{fig:system_desc_nested}).

Following the approach of Mantri \textit{et al.}~\cite{mantri2025comparing}, we explicitly track the \emph{full probability distribution} of the number of available Bell pairs at each level. This enables a unified treatment of multi-success multiplexing, distillation, and nested swapping within a recursive framework (see Appendix~\ref{appendix:recursive_calcs} for details).

\subsection{Noise and Error Models}

\paragraph{Initial state model.}
Generated elementary links are modeled as Bell-diagonal states
\begin{equation}
\rho = a\ket{\phi^+}\bra{\phi^+}
+ b\ket{\phi^-}\bra{\phi^-}
+ c\ket{\psi^+}\bra{\psi^+}
+ d\ket{\psi^-}\bra{\psi^-},
\end{equation}
where $\ket{\phi^{\pm}} = (1/\sqrt2)(\ket{00} \pm \ket{11})$, and $\ket{\psi^{\pm}} = (1/\sqrt2)(\ket{01} \pm \ket{10})$, represent the four canonical Bell states, with fidelity being defined on state $\ket{\phi^+}$, i.e., $F = a$, $0< a \le 1$. The initial fidelity is determined by local gate errors and state-preparation imperfections. Under the depolarizing noise model used in~\cite{muralidharan2016optimal, mantri2025comparing}, this yields
\begin{align*}
    F_0 \approx 1 - \frac{5}{4}\epsilon_G,
\end{align*}
where $\epsilon_G$ denotes the two-qubit gate error probability.

\paragraph{Gate and measurement noise.}
Two-qubit gates are modeled via a depolarizing channel with error probability $\epsilon_G$:
\begin{equation}
\mathcal{N}_{\tilde{U}}(\rho)
= (1-\epsilon_G) U \rho U^\dagger
+ \frac{\epsilon_G}{4} \mathrm{Tr}_{ij}[\rho]\otimes I.
\end{equation}
Measurement errors occur with probability $\xi = \epsilon_G/4$, consistent with the depolarizing noise model used in prior work~\cite{muralidharan2016optimal, mantri2025comparing}.

\paragraph{Memory decoherence.}
We include pure dephasing noise characterized by coherence time $T_2$, updating the Bell-diagonal coefficients according to $  \Lambda_{\mathrm{dec}} = (1 + e^{-2t/T_2})/2$,
where $t$ is the storage duration. Relaxation ($T_1$) processes are neglected, consistent with the dominant noise model in long-lived quantum memories in this regime \cite{muralidharan2016optimal, mantri2025comparing}. 

\subsection{Nested Swapping Strategy}

We employ a deterministic nested swapping schedule, following the ``Innsbruck protocol'' introduced by D\"ur \textit{et al.}~\cite{dur1999quantum} and further analyzed in subsequent studies~\cite{hartmann2007role}, dividing the network into $N = 2^n$ elementary links. Swaps are performed recursively to double link length at each nesting level (See Fig.~\ref{fig:system_desc_nested}).
Deterministic swapping simplifies the recursive probability analysis and avoids additional stochastic branching. While swapping extends distance, it causes fidelity decay that grows with nesting depth. Therefore, swapping is interleaved with optional distillation steps to maintain end-to-end link.

\subsection{Distillation Model and Sequencing}

We adopt a probabilistic DEJMPS~\cite{deutsch1996quantum, muralidharan2016optimal, mantri2025comparing} ($2\rightarrow1$) distillation protocol. For two input Bell-diagonal states $(a_1,b_1,c_1,d_1)$ and $(a_2,b_2,c_2,d_2)$, the output state and success probability are computed using the standard noisy DEJMPS expressions. To maintain tractability and align with high-fidelity operating regimes, we allow at most one round of distillation per nesting level, aimed towards regimes where high initial fidelities limit the benefit of repeated rounds.

Distillation decisions are not reactive to real-time outcomes but are pre-computed and fixed across bursts. We consider a distillation strategy based on a fidelity-threshold ($F_{\mathrm{th}}$ rule). In this rule, we distill at level $i$ if fidelity at nesting level $i$ (denoted $F_i$) falls below a pre-determined fidelity threshold, $F_{\mathrm{th}}$.

\section{Methods}\label{sec:Methods}

In this section we describe the simulation framework used to evaluate the performance of the repeater networks. Sec.~\ref{sec:recursive} introduces the recursive formulation used to track the probability distribution of the number of Bell pairs delivered end-to-end, with additional details provided in Appendix~\ref{appendix:recursive_calcs}. Sec.~\ref{sec:termination} provides a note on protocol termination, and Sec.~\ref{sec:summary} provides a summary of the major assumptions. Sec.~\ref{sec:memory_tofiber_coupling} describes the model used to compute the memory-to-fiber coupling efficiency.

\subsection{Recursive Probability Tracking}\label{sec:recursive}
To analyze the performance of multiplexed two-way protocols, we adopt a recursive probability-tracking framework following Ref.~\cite{mantri2025comparing}, specialized to the architectural regime considered in this work.

Let $Y_i$ denote the number of Bell pairs available at nesting level $i$. The distribution $p_{i,k} = \Pr(Y_i = k)$ is computed recursively by combining probabilistic elementary link generation, probabilistic distillation, deterministic swapping, and a pre-defined termination threshold.

Tracking the full distribution is essential because the protocol evolution is governed by nonlinear, conditional operations---including both the pairing of Bell pairs across adjacent logical links and interleaved distillation steps. In particular, swapping is bottlenecked by $\min(k_{\text{left}}, k_{\text{right}})$, while distillation involves thresholding and consumption of multiple pairs. The expectation alone does not determine quantities such as $\mathbb{E}[\min(k_{\text{left}}, k_{\text{right}})]$ or the probability of successful distillation, which depend on the variance and tail behavior of the distributions. This recursive formulation allows computation of the expected number of surviving Bell pairs, reset probabilities, and expected secret-key yield per burst, as detailed in the Appendix~\ref{appendix:recursive_calcs}. 

\subsection{Termination Strategy}\label{sec:termination}

If at any nesting level $i$ the number of available Bell pairs is at most $1$, the protocol may terminate or proceed without further distillation. We optimize operation so as to minimize classical communication latency owing to protocol termination---see~\cite{mantri2025comparing} for details.

\subsection{Evaluation Metrics} \label{sec:eval_metric}

The repeater architectures considered in this work may support a range of applications, including quantum key distribution, distributed quantum sensing, and distributed quantum computation. Our objective is therefore not to optimize the network for a particular application-layer protocol. Instead, we seek a common end-to-end metric that captures both the quantity of entanglement delivered and the quality of the resulting states.

For this purpose, we use the asymptotic secret-key yield of the entanglement-based BBM92 protocol~\cite{bennett1992quantum} as an operational benchmark. Proposed by Bennett \textit{et al.} in 1992, the BBM92 provides a convenient mapping from the number and quality of the delivered Bell pairs to a single scalar quantity and is directly applicable to the Bell-diagonal states produced by the repeater protocol. The resulting secret-key rate should therefore be interpreted primarily as a normalized measure of usable end-to-end entanglement, rather than as an assumption that BBM92 is the only intended application of the architectures considered here.

For an end-to-end Bell-diagonal state

\begin{align*}
    \rho_n &=
            a_n \ket{\phi^+}\!\bra{\phi^+}
            +
            b_n \ket{\phi^-}\!\bra{\phi^-}
            +
            c_n \ket{\psi^+}\!\bra{\psi^+}
            +
            d_n \ket{\psi^-}\!\bra{\psi^-},
\end{align*}

the quantum bit error rates in the $Z$ and $X$ measurement bases are
\begin{align*}
    Q_Z = c_n + d_n, \qquad Q_X = b_n + d_n.
\end{align*}
The corresponding asymptotic secure-key fraction per successfully delivered Bell pair is
\begin{align*}
    r_{\mathrm{secure}} =\max \left\{0,\,1-h_2(Q_X)-h_2(Q_Z)\right\},
\end{align*}
where
\begin{align*}
    h_2(x) = -x\log_2 x - (1-x)\log_2(1-x)
\end{align*}
is the binary entropy function.

Let $\mathbb{E}[Y_n]$ denote the expected number of end-to-end Bell pairs delivered in one protocol burst, and let $M$ denote the number of multiplexed elementary-link channels used in that burst. We define the secret-key rate per channel use per shot as
\begin{align*}
    \mathrm{SKR/PCU} = \frac{\mathbb{E}[Y_n]}{M} r_{\mathrm{secure}}.
\end{align*}
This metric decreases when either fewer Bell pairs are delivered or the delivered states accumulate excessive errors. It therefore captures, within a single normalized quantity, the rate--quality tradeoff induced by transmission loss, memory decoherence, imperfect local operations, entanglement swapping, and distillation.

\subsection{Summary of Major Assumptions}\label{sec:summary}

\begin{itemize}[itemsep = 0em]
    \item High-fidelity elementary links ($F \approx 1 - O(\epsilon_G)$).
    \item Two-photon, meet-in-the-middle elementary-link entanglement generation.
    \item Static, pre-computed distillation schedule based on the $F_{\mathrm{th}}$ rule, under which distillation is performed at nesting level $i$ whenever the predicted fidelity $F_i$ falls below the prescribed threshold $F_{\mathrm{th}}$.
    \item Long-lived memories with $T_2$ exceeding classical signaling times.
    \item Deterministic swapping.
    \item Probabilistic DEJMPS distillation.
    \item Single distillation round per level.
    \item Full probability-distribution tracking for multiplexed channels.
\end{itemize}

These assumptions reflect an optimistic but experimentally motivated parameter regime, consistent with the architectural comparison framework in \cite{mantri2025comparing}. Fig.~\ref{fig:loss} compares the propagation loss of SMF and HCF, specifically DNANF~\cite{Poletti14nanf, NumkamFokoua23hcfreview, Petrovich2025}. Based on this we use $L^\mathrm{SMF}_\mathrm{att}(1550) = 28.95$ km, $L^\mathrm{HCF}_\mathrm{att}(780) = 24.127$ km,  $L^\mathrm{HCF}_\mathrm{att}(1550) = 78.96$ km, corresponding to values reported in Fokoua \textit{et al.} 2023 \cite{NumkamFokoua23hcfreview}. Equivalently, these correspond to $0.150$, $0.180$, and $0.055$~dB/km, respectively. The SMF attenuation represents state-of-the-art performance in pure silica core fiber rather than typical deployed SMF ($\sim0.20$~dB/km), and is therefore conservative here. The HCF losses are taken from the projected loss curves~\cite{NumkamFokoua23hcfreview}, while the lowest measured km-scale attenuations are $0.091$\,dB/km at 1550\,nm over 15\,km~\cite{Petrovich2025} and $0.33$\,dB/km at 850\,nm over 10.9\,km~\cite{Adamu:24}.

We assume a memory-native emission wavelength of $780\,\mathrm{nm}$, representative of near-$800\,\mathrm{nm}$ quantum-memory emission wavelengths. While this wavelength is particularly relevant for rubidium-based memory systems~\cite{Hsiao2018}, the analysis is intended to represent the broader class of near-$800\,\mathrm{nm}$ memory emitters. Such memory-native wavelengths can, in principle, be transmitted directly when supported by the transmission medium and detection architecture, avoiding frequency-conversion losses. For telecom-band transmission, we include two-stage aggregate quantum frequency-conversion losses, where photons are converted from $780\,\mathrm{nm}$ to $1550\,\mathrm{nm}$ for transmission and subsequently converted back to the detector operating wavelength ($\sim800\,\mathrm{nm}$ for silicon avalanche photodiodes (APDs)) using a meet-in-the-middle architecture~\cite{jones2016design}. We model optical transmission as a pure-loss channel characterized by the attenuation lengths above and neglect other impairments such as dispersion, phase noise, or Raman scattering at the level of this analysis. As an additional analysis, we also consider direct telecom-band detection using SNSPDs, in which photons are converted once from the memory wavelength to $1550\,\mathrm{nm}$ and detected directly at the telecom wavelength.

\vspace{10pt}

\subsection{Memory-to-fiber coupling efficiency}\label{sec:memory_tofiber_coupling}

\begin{wrapfigure}{R}{0.45\textwidth} 
  \centering
  \includegraphics[width=\linewidth]{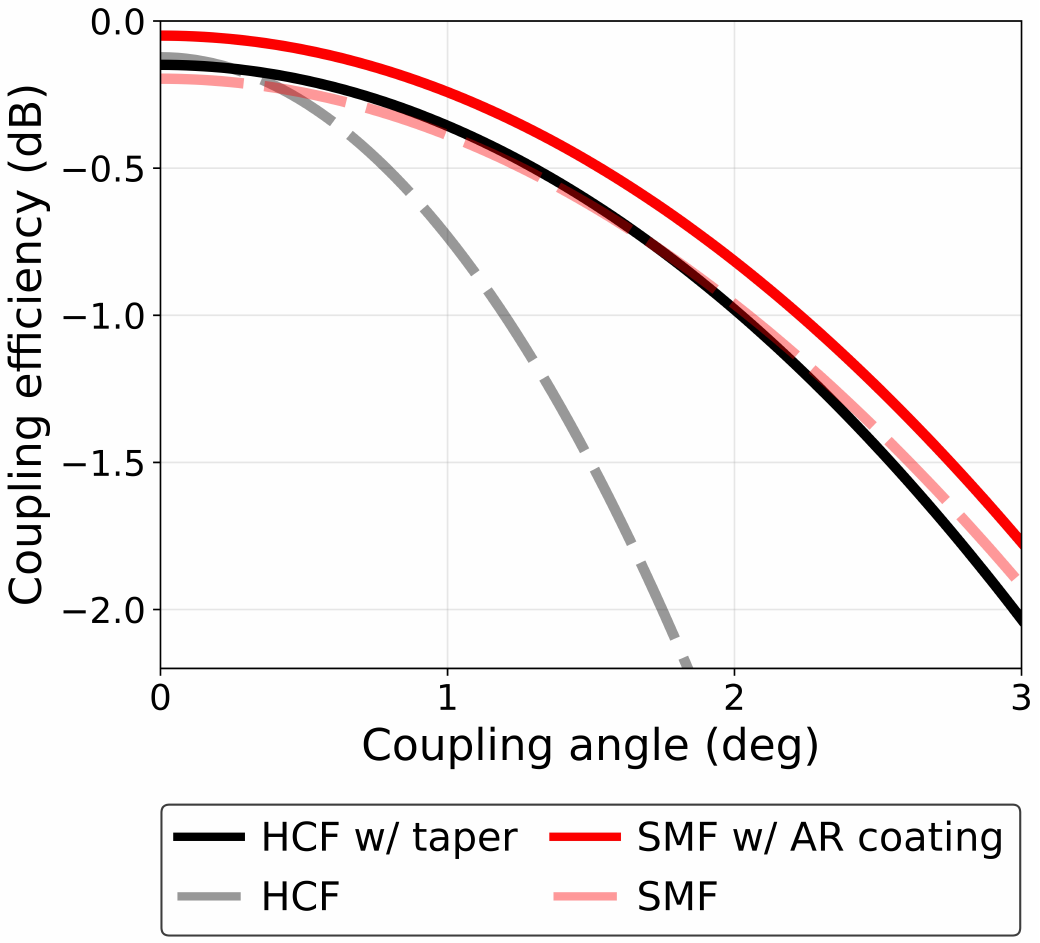} 
  \caption{\textbf{Angle-dependent mode-match coupling efficiency $\eta_{\mathrm{opt}}(\theta)$ for conventional SMF at 1550\,nm and DNANF designs at 1550\,nm and 780\,nm.} Angular misalignment is modeled as a tilt by $\theta$ in the $x$--$z$ plane. The non-tapered DNANF shows reduced angle tolerance due to its larger core, while adiabatically tapered end facets improve tolerance and approach SMF performance.}
  \label{fig:angle-comp}
  \vspace{-0.1cm}
\end{wrapfigure}

Coupling efficiency plays a critical role in determining network performance. As seen from Eq.~\eqref{eq:p_succ_elem}, the elementary entanglement generation probability scales quadratically with $\eta_c$, since successful entanglement generation requires two photons to be transmitted and detected. Consequently, even modest improvements in coupling efficiency can lead to order-of-magnitude changes in end-to-end key rates and optimal repeater spacing. This sensitivity has been highlighted in previous studies of both one-way and two-way repeater architectures~\cite{muralidharan2016optimal,mantri2025comparing}.

In practice, the dominant technology-dependent contribution arises from the memory-to-fiber coupling efficiency $\eta_{\mathrm{coupling, memory\to fiber}}$. This coupling depends strongly on the mode structure and confinement properties of the transmission medium. Since mechanisms of light guidance differ fundamentally between hollow-core fibers and conventional silica single-mode fibers, the achievable coupling efficiencies can differ significantly between the two technologies. In particular, the larger core of low-loss DNANF ($D\sim30 ~\mu$m) requires a substantially larger incident mode-field diameter than conventional SMF ($\sim 10~\mu$m). In the following sections we evaluate the resulting architectural performance differences under realistic coupling and conversion assumptions.

\paragraph{Ideal Gaussian beam coupling:} 
To compare the memory-to-fiber coupling efficiencies, we consider coupling of a free space Gaussian beam to the traditional SMF and DNANF cross-sections. In general, we compute the power coupling efficiency $\eta$ using the vector overlap integral between the incident Gaussian field ($\mathbf{E}_{\text{inc}}, \mathbf{H}_{\text{inc}}$) and the computed fiber mode ($\mathbf{E}_{\text{fib}}, \mathbf{H}_{\text{fib}}$), given by
\begin{equation}
    \eta(w) = \frac{\left| \frac{1}{2} \iint \left( \mathbf{E}_{\text{inc}} \times \mathbf{H}_{\text{fib}}^* + \mathbf{E}_{\text{fib}}^* \times \mathbf{H}_{\text{inc}} \right) \cdot \hat{\mathbf{z}} \, dA \right|^2}{4 P_{\text{inc}} P_{\text{fib}}}\label{eq:coupling_vector}
\end{equation}
where $^*$ denotes the component-wise complex conjugate of the vector fields and $P=\iint\frac{1}{2}\mathrm{Re}\left(\mathbf{E}\times \mathbf{H}^*\right)\cdot\hat{z}dA$ represents the optical power carried by the respective fields, and where $\mathrm{Re}(\cdot)$ denotes the real value component. 
For a traditional single-mode fiber with core radius $a$, the weakly guiding approximation allows us to use a scalar transverse field $\psi_\text{fib}(r)$, reducing \eqref{eq:coupling_vector} to the normalized scalar overlap in polar coordinates.  Using the Helmholtz equation \cite[Eq.~(10.2-6)]{SalehTeich2019}, we have
\begin{equation}
    \psi_{\text{fib}}(r) = 
    \begin{cases} 
      J_0\left(U \frac{r}{a}\right) & r \le a \quad (\text{Core}) \\
      \frac{J_0(U)}{K_0(W)} K_0\left(W \frac{r}{a}\right) & r > a \quad (\text{Cladding})
   \end{cases}
\end{equation}
where $J_n$ (resp. $K_n$) is the (resp. modified) Bessel function of first (resp. second) kind, with order $n$. We define the normalized transverse field parameters 
\begin{align}
    U &\;:=\; a\sqrt{n_1^2k_0^2-\beta^2},\label{eq:U}\\
    W &\;:=\; \; a\sqrt{\beta^2-n_2^2k_0^2},\label{eq:W}
\end{align}
where $k_0=2\pi/\lambda$ is the free-space wavenumber, $n_1$ and $n_2$ are the core and cladding refractive indices, and $\beta$ is the mode propagation constant. Here, $U$ is the normalized transverse wavenumber in the core, and $W$ is the normalized decay constant in the cladding such that guided modes require $U^2>0$ and $W^2>0$, and hence $n_2k_0<\beta<n_1k_0$. The normalized frequency is given by
\begin{align}
    V & \;:=\; a k_0\sqrt{n_1^2-n_2^2}\; =\; \frac{2\pi a}{\lambda}\mathrm{NA}, \label{eq:V}
\end{align}
where $\mathrm{NA}=\sqrt{n_1^2-n_2^2}$ is the numerical aperture of the fiber. $V$ governs the number of modes in the fiber at a particular wavelength, and is completely determined by the fiber geometry $a$ and index contrast. Note that \eqref{eq:U}-\eqref{eq:V} imply $V^2=U^2+W^2$.
The transverse parameters $U$ and $W$ relate to the normalized frequency $V$ via the characteristic equation \cite[Eq.~(10.2-14)]{SalehTeich2019}
\begin{align}
\frac{U J_1(U)}{J_0(U)} &= \frac{W K_1(W)}{K_0(W)}.
\end{align}
For single-mode operation, we have $V < 2.405$ at the first root of $J_0$. For near-cutoff single-mode fiber, we have $U \approx 1.645$ and $W \approx 1.754$.
Using a Gaussian beam at the waist ($z=0$) with radius $w$ defined as $\psi_\text{inc}(r;w) = \exp\left(-\frac{r^2}{w^2}\right),$ we compute the ultimate limit of power coupling efficiency $\eta$ as \cite{marcuse77loss} 
\begin{equation}
    \eta(w) = \frac{\left| \int_0^\infty \psi_\text{fib}(r) \psi_\text{inc}(r; w) 
    \, r \, dr \right|^2}{\left(\int_0^\infty |\psi_\text{fib}(r)|^2 \, r \, dr\right) 
    \left(\int_0^\infty |\psi_\text{inc}(r; w)|^2 
    \, r \, dr\right)}.
\end{equation}
We maximize $\eta$ with respect to the Gaussian width $w$ numerically for $V=2.405$, which yields 
\begin{align}
\eta_{\text{opt}} &=\eta(w_{\text{opt}})\approx 0.997\text{, with}\\
    w_{\text{opt}} &\approx 1.09 a.
\end{align}

We capture the complex vector modes of the DNANF fiber geometry \cite{Petrovich2025} via Finite-Difference Eigenmode (FDE) solver (Tidy3D) with PML boundaries. The simulated geometry consists of a $D = 29.35\, \mu\text{m}$ air core surrounded by a double-nested ring of $N=5$ silica capillaries with thickness $t = 500$\,nm.
Unlike the scalar approximation sufficient for step-index fibers, the microstructured cladding necessitates a full vector analysis.  Sweeping the beam waist $w$ allows us to identify the coupling limit imposed by the mode shape mismatch of the 5-tube geometry, as in \cite{Zuba2023CouplingEfficiency}. We find
\begin{align}
\eta_{\text{opt}} &=\eta(w_{\text{opt}})\approx0.98  \text{, with}\\
    w_{\text{opt}} &\approx 0.37D.
\end{align}

Integrated photonic sources interface to fiber through several mechanisms, including inverse-taper edge couplers, grating couplers, and adiabatic tapered-fiber couplers. Where the interface crosses free space, the design objective is a facet mode closely matched to the target fiber, and the governing requirement is the target mode-field diameter: $\approx9~\mu$m for SMF versus $\approx 22~\mu$m for DNANF. Residual departures from a Gaussian facet mode reduce the absolute overlap for both fiber types and enter the SMF/HCF comparison as a common factor.

\noindent{\bf Coupling non-idealities:}
We consider the impact of angle mismatch and air-to-fiber reflections.
For both fiber types, we compute the angular tolerance by modeling a rotation of the incident beam by an angle $\theta$ in the $x$--$z$ plane, which introduces a transverse phase ramp across the fiber facet:
\begin{equation}
    \mathbf{E}_\text{inc}(x,y;\theta)=\mathbf{E}_\text{inc}(x,y;0)\exp\!\left(i k_0 \sin\theta \, x\right)
\end{equation}
Applying a similar transformation for $\mathbf{H}_\text{inc}$ and substitution into \eqref{eq:coupling_vector} yields our angle-dependent mode-match coupling coefficient $\eta_\text{opt}(\theta)$.
In addition to mode mismatch, an uncoated SMF facet introduces Fresnel reflection at the air--silica interface. For near-normal incidence from air ($n_0\simeq 1$) into silica ($n_1\simeq 1.45$), the power reflectance is
\begin{equation}
    R=\left(\frac{n_1-n_0}{n_1+n_0}\right)^2,
\end{equation}
so that the power transmission into the fiber is $T=1-R$. These reflection losses can be mitigated through the use of an anti-reflective coating. For the DNANF design, the mode is mostly confined in air, resulting in negligible losses due to reflections.

We find that while the DNANF design has slightly better incident coupling due to lack of Fresnel reflections, it is less tolerant to angle mismatch than traditional SMF at both wavelengths. This is due to the much larger core diameter of the DNANF ($\sim 30\,\mu\mathrm{m}$) compared to traditional SMF ($\sim 10\,\mu\mathrm{m}$). To improve angle tolerance, we consider an adiabatically tapered DNANF design with constant capillary thickness and end-facet core diameter of $\sim18\,\mu\mathrm{m}$ at 1550\,nm and $\sim9\,\mu\mathrm{m}$ at 780\,nm. To model these, we scale the 5-tube DNANF geometry by 0.6 and 0.3, respectively, and compute the angle-dependent coupling efficiency $\eta_\text{opt}(\theta)$.  

Angle tolerance is shown in Fig.~\ref{fig:angle-comp}. We find that the tapered design has similar coupling efficiency to AR-coated SMF for both wavelengths across moderate angle mismatch.  In our simulations, we use input-output coupling efficiencies of 0.79 for HCF and 0.83 for SMF, corresponding to an angle tolerance of 0.025 radians ($\sim1.4^\circ$).

Our analysis here assumes light is launched directly into the fiber facet. For an integrated source this is demanding, since the $\sim 22~\mu$m target MFD exceeds what conventional edge couplers deliver, and no adiabatically tapered nested-antiresonant end facet has yet been experimentally demonstrated. Instead, a deployable alternative can avoid direct chip-to-HCF coupling entirely by employing a high-efficiency edge-coupler to a short high-NA solid core fiber pigtail, followed by a graded-index mode-field adapter to the HCF. Back-reflection at the reintroduced air-glass interface can be mitigated by anti-reflective coating and angle-cleaved splicing~\cite{Suslov2021, Zhang26}. Both stages have been demonstrated at 1550~nm. Adiabatic silicon nitride inverse-taper edge couplers have achieved $\approx0.966$ coupling efficiency per facet to high-NA fiber at 1550 nm~\cite{Zhu16,Avdeev25} and SMF-HCF interconnection with throughput efficiency of $0.973$~\cite{Zhong24}. The combined efficiency $\eta\approx0.94$ exceeds both values assumed above. Notably, the added SMF-HCF interconnection loss $(\approx1-0.973)$ is smaller than the per-facet coupling penalty already assigned to HCF relative to SMF ($\approx1-0.952$), so routing through a solid-core pigtail costs less than the direct-coupling disadvantage it removes.

These results suggest that HCFs can be incorporated into otherwise SMF-based links---e.g., using HCF for long-haul transmission and SMF for local distribution and pig-tailing---without requiring end-to-end HCF infrastructure.


\section{Results}\label{sec:Results}

\begin{figure}[!htb]
    \centering
    \begin{subfigure}{0.95\linewidth}
    \caption{Perfect Conversion Efficiency}
    \includegraphics[width=\linewidth,
                         trim=0cm 2cm 0cm 2cm, clip]{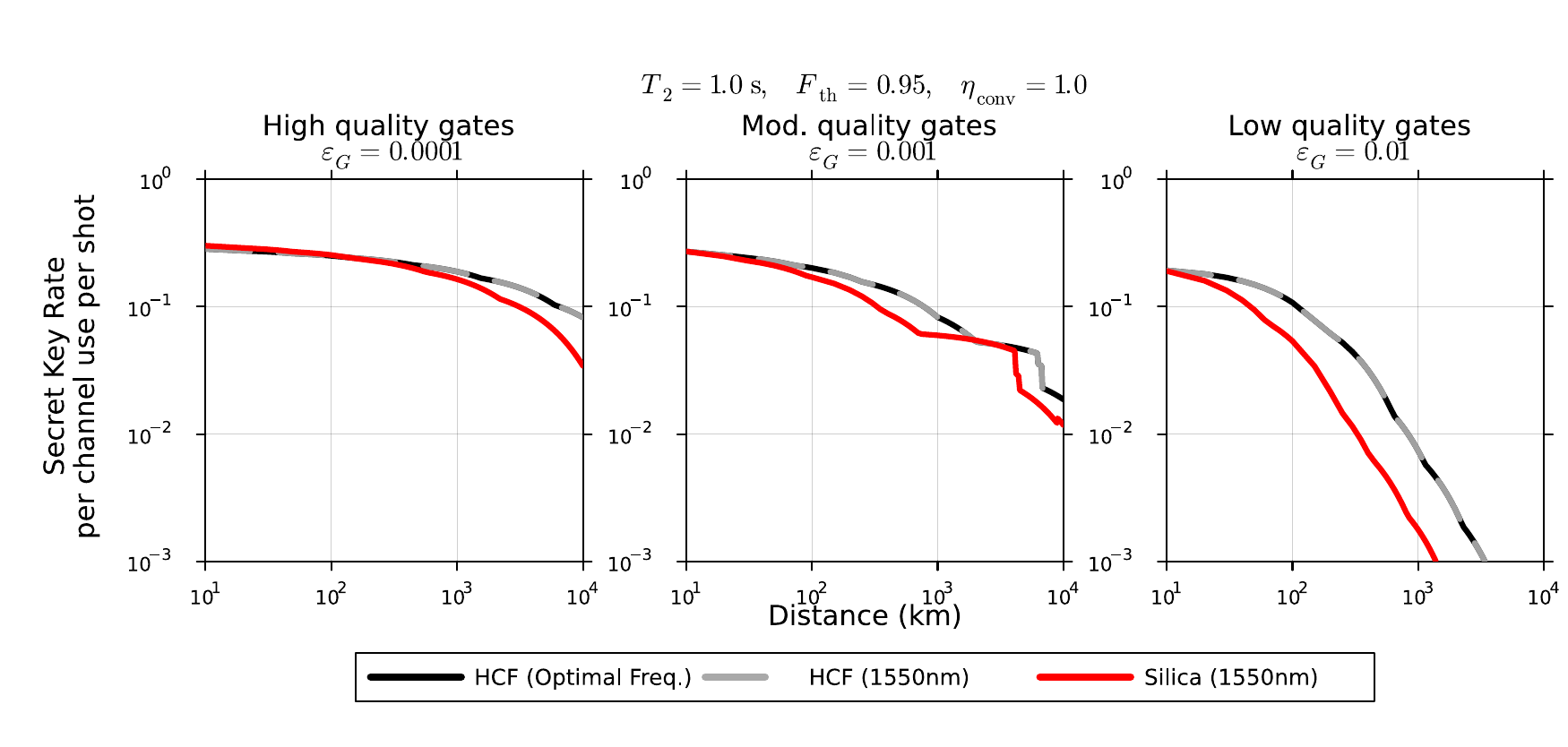}
\end{subfigure}
    \begin{subfigure}{0.95\linewidth}
    \caption{Moderate Conversion Efficiency}
    \includegraphics[width=\linewidth,
                         trim=0cm 0cm 0cm 2cm, clip]{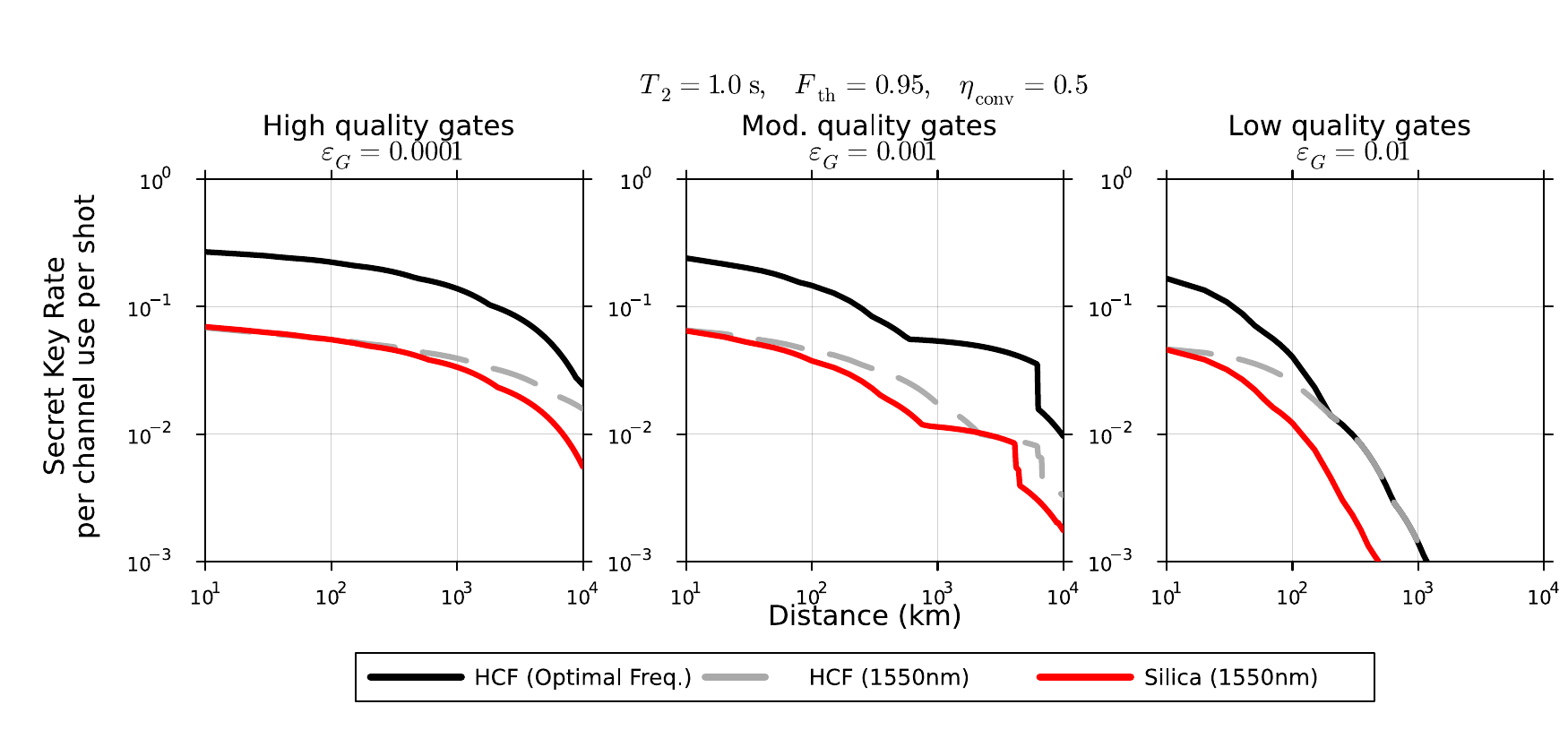}
\end{subfigure}
    \caption{
    \textbf{Performance comparison between HCF and silica SMF quantum repeater architectures.} The secret-key rate per channel use per shot (SKR/PCU) is shown as a function of the total end-to-end distance between Alice and Bob. Columns correspond to gate-quality regimes (high quality gates -- $\epsilon_G = 10^{-4}$, medium quality -- $\epsilon_G = 10^{-3}$, and low quality -- $\epsilon_G = 10^{-2}$), illustrating how network performance degrades as local gate operations become noisier. Rows correspond to the assumed two-stage aggregate frequency conversion efficiency: perfect conversion ($\eta_\mathrm{conv}=1.0$, top row) and moderate conversion ($\eta_\mathrm{conv}=0.5$, bottom row).
    Black curves denote HCF operating at the \emph{optimal wavelength}, defined as the choice between the memory-native emission wavelength ($\sim$780\,nm) and telecom wavelength (1550\,nm) that maximizes the elementary entanglement generation probability for the link. Gray dashed curves denote HCF operation constrained to telecom wavelength (1550\,nm), while red curves denote conventional silica SMF operation at 1550\,nm. The distillation threshold is fixed at $F_{\mathrm{th}} = 0.95$ in all simulations.
    Across all gate-quality regimes, HCF achieves higher secret-key rates than SMF, with the advantage becoming particularly pronounced at long distances and when frequency conversion efficiency is limited. Even when restricted to telecom operation at 1550\,nm, HCF maintains performance comparable to or better than silica SMF, highlighting the architectural advantages of HCF links in long-distance repeater networks.
    }
\label{fig:perf_eval}
\end{figure}

\begin{figure}[!htb]
    \centering
    \includegraphics[width=\linewidth,
                         trim=0cm 0cm 0cm 2.1cm, clip]{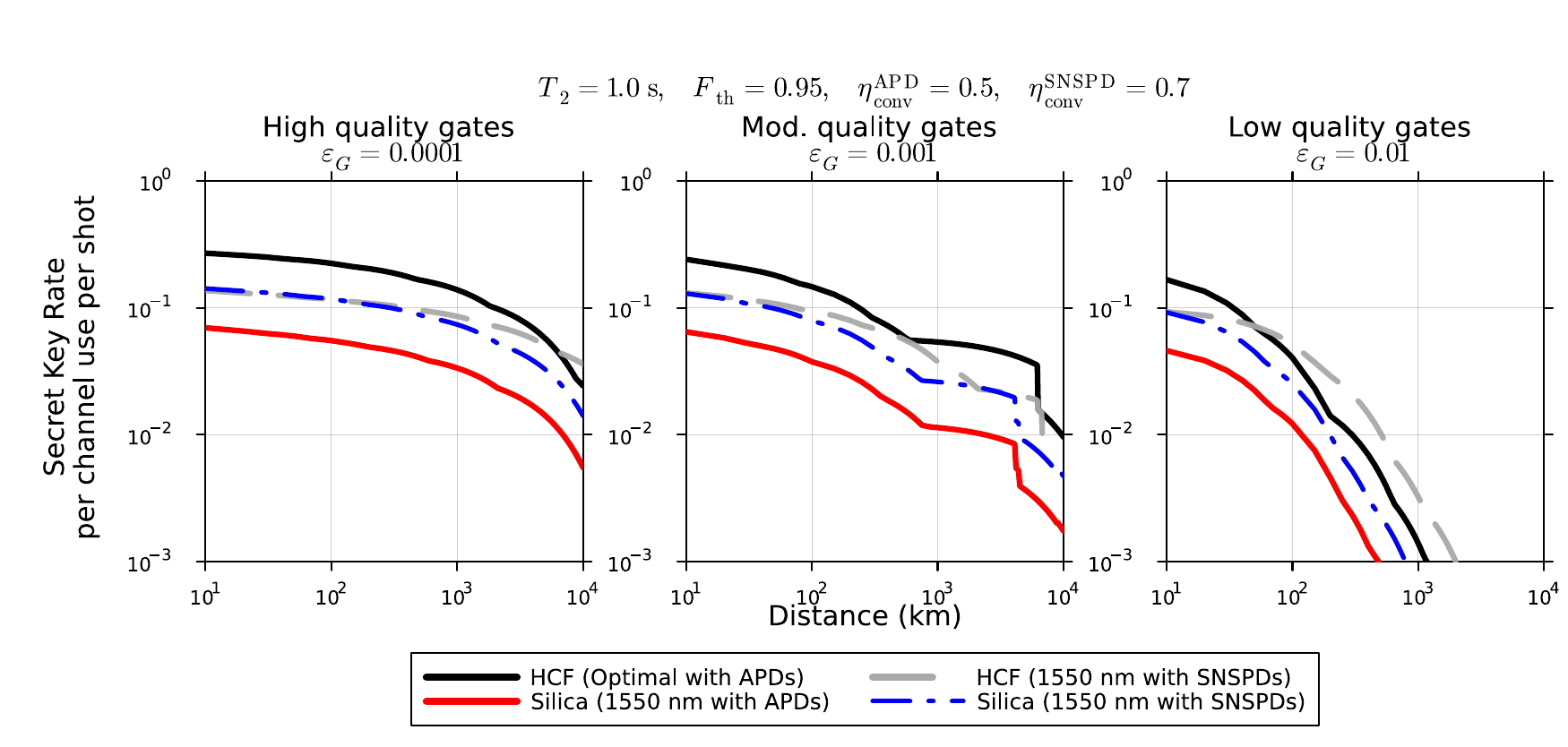}
\caption{\textbf{Comparison of architectures using silicon avalanche photodiode detectors (APD) and superconducting nanowire single-photon detectors (SNSPDs).}
Secret-key rate per channel use per shot as a function of total end-to-end distance for three two-qubit gate-error probabilities. Black curves denote HCF operated using the original APD-based architecture, with the transmission wavelength chosen adaptively between direct memory-native transmission at approximately $780~\mathrm{nm}$ and telecom transmission at $1550~\mathrm{nm}$ followed by conversion to the APD detection wavelength. Red curves denote silica SMF using telecom transmission with APD detection. Gray dashed and blue dash-dotted curves denote HCF and silica SMF, respectively, using conversion from the memory wavelength to $1550~\mathrm{nm}$ followed by direct telecom-band SNSPD detection, thereby eliminating the second frequency-conversion stage. The aggregate conversion efficiency is assumed to be $\eta_{\mathrm{conv}}=0.7$ for a single down-conversion stage ($780\rightarrow1550~\mathrm{nm}$) and $\eta_{\mathrm{conv}}=0.5$ for the combined down- and up-conversion process ($780\rightarrow1550\rightarrow800~\mathrm{nm}$). We assume $T_2=1~\mathrm{s}$, $F_{\mathrm{th}}=0.95$, and $M=1024$. Direct telecom SNSPD detection improves the performance of both fiber technologies by removing the additional conversion loss. Nevertheless, even when restricted to APD-based detection, optimally operated HCF outperforms silica SMF with direct telecom SNSPD detection across all considered hardware regimes. Allowing HCF to additionally exploit direct telecom SNSPD detection further improves its performance, particularly in regimes with lower-quality local operations.}
\label{fig:snsdp_comparison}
\end{figure}

\begin{figure}[!htb]
\centering
        \begin{subfigure}{0.4\linewidth}
            \includegraphics[width=\linewidth]{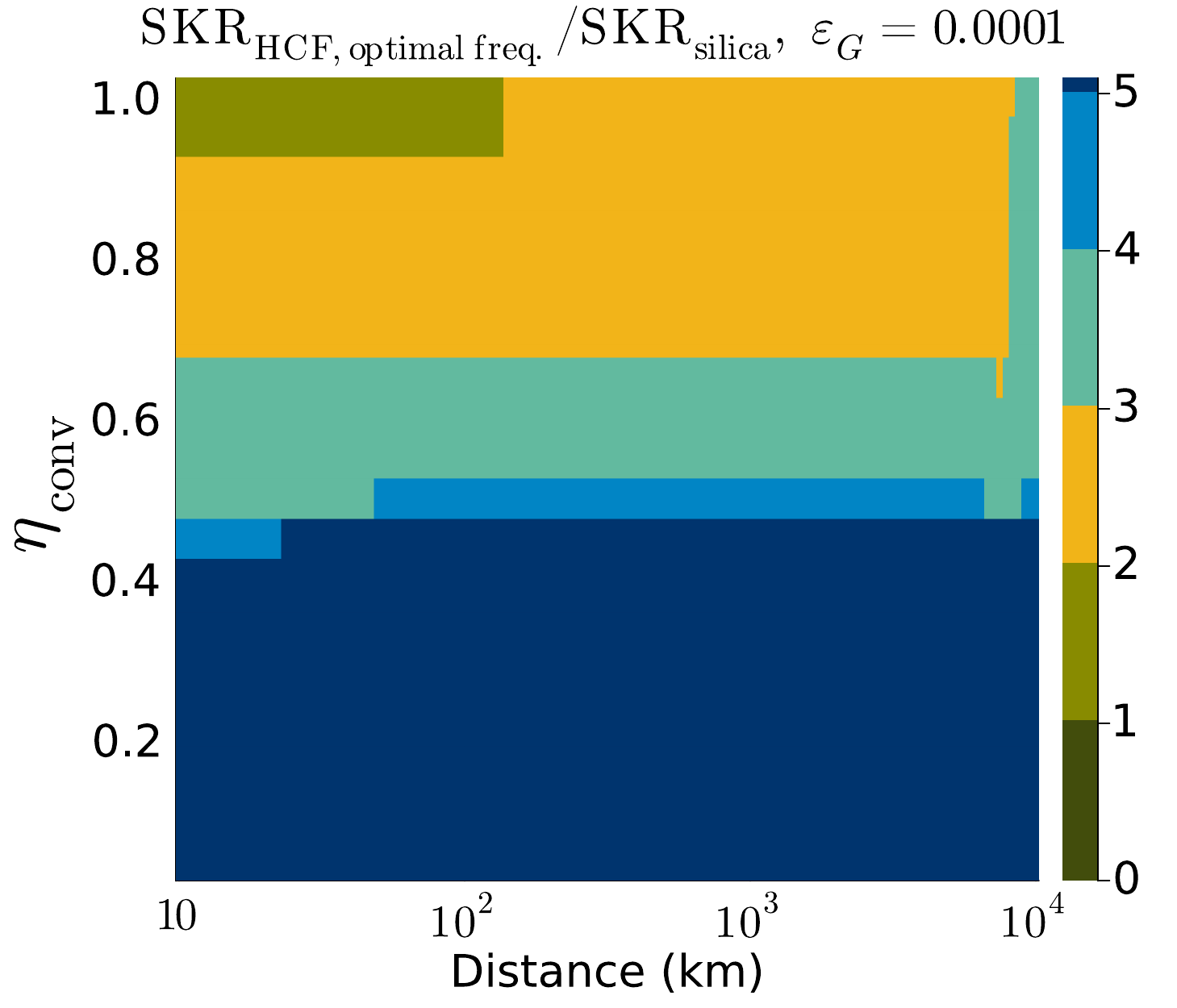}
        \end{subfigure}
        \begin{subfigure}{0.4\linewidth}
            \includegraphics[width=\linewidth]{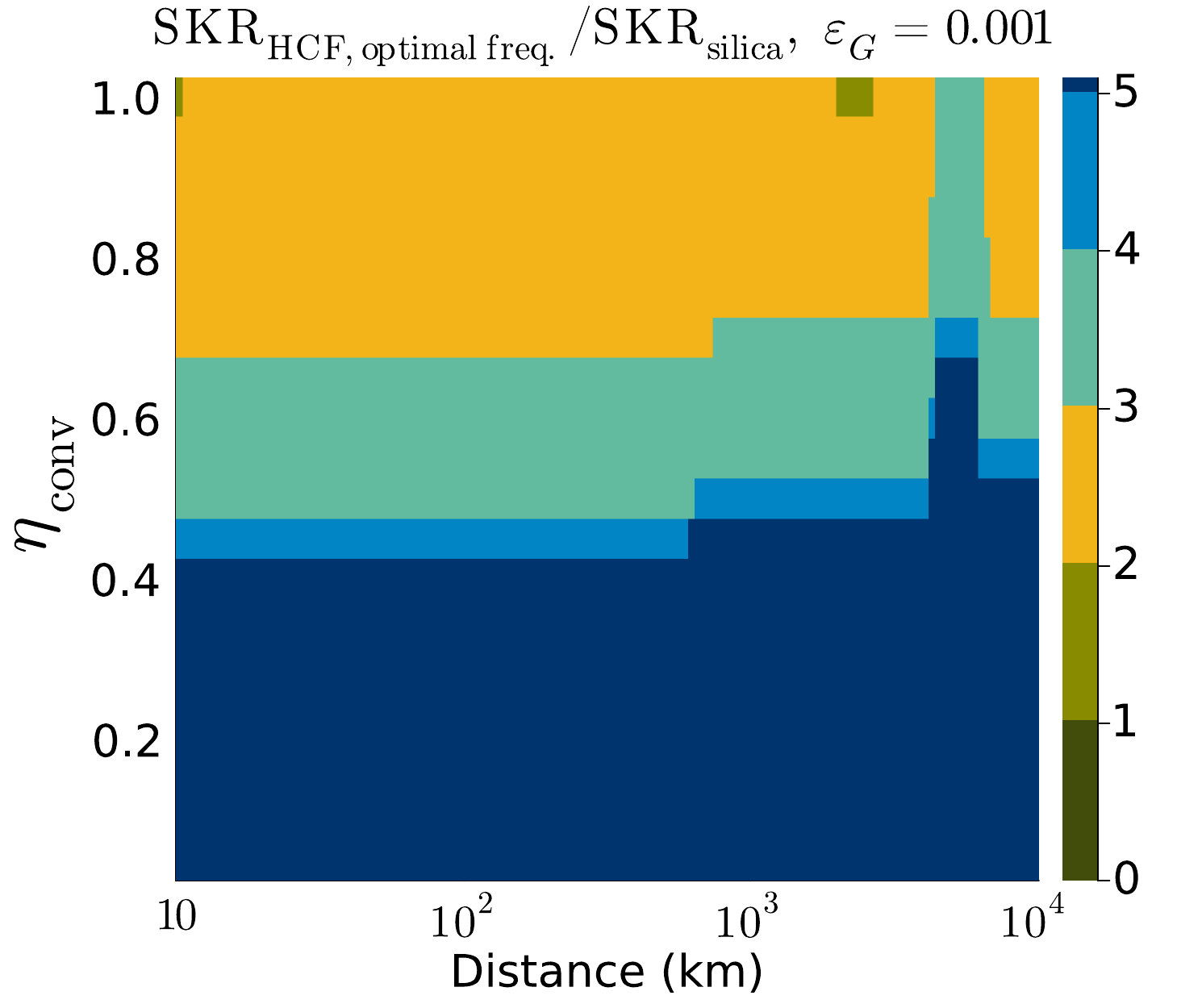}
        \end{subfigure}
    \caption{\textbf{Ratio $\mathrm{\textbf{SKR}}_{\mathrm{\textbf{HCF,opt}}}/\mathrm{\textbf{SKR}}_{\mathrm{\textbf{silica}}}$ as a function of total distance and conversion efficiency.}
    The left and right panels correspond to gate error rates $\epsilon_G=10^{-4}$ and $\epsilon_G=10^{-3}$, respectively. Colors denote binned values of the ratio, with larger values indicating stronger advantage for adaptive HCF. 
    The region $\mathrm{SKR}_{\mathrm{HCF,opt}}/\mathrm{SKR}_{\mathrm{silica}}<1$, where silica would outperform HCF, is absent across the plotted range, indicating that adaptive HCF matches or exceeds silica performance throughout nearly the entire parameter space. In particular, at long distances or limited conversion efficiency, HCF greatly outperforms SMF.
    }
    \label{fig:skr_ratio_conv_eff}
\end{figure}

This section compares HCF- and silica-SMF-based repeater architectures using the evaluation framework introduced in Sec.~\ref{sec:eval_metric}. Our principal performance metric is the secret-key rate per channel use per shot, which combines the expected number of end-to-end Bell pairs delivered by the protocol with the asymptotic secure-key fraction supported by their final error rates. We use this quantity as a normalized operational measure of usable entanglement, rather than to restrict the architectural comparison to a particular QKD deployment.

We complement this performance metric with architectural and resource-oriented measures, including the optimal inter-repeater spacing, and the number of repeater operations required per delivered secret key. The analysis focuses on three aspects: (i) performance under gate, memory, and frequency-conversion noise; (ii) robustness to hardware-related inefficiencies captured by $\eta_{\mathrm{hardware}}$; and (iii) the operational resources required to sustain a given secret-key throughput. These metrics characterize both the achievable end-to-end performance and the architectural resource requirements of HCF- and silica-based repeater networks.

\subsection{Effect of operations and memory noise on performance}\label{sec:results_gatenoise}
\begin{figure}[!t]
\centering
        \includegraphics[width=0.95\linewidth,
                         trim=0cm 0cm 0cm 2cm, clip]{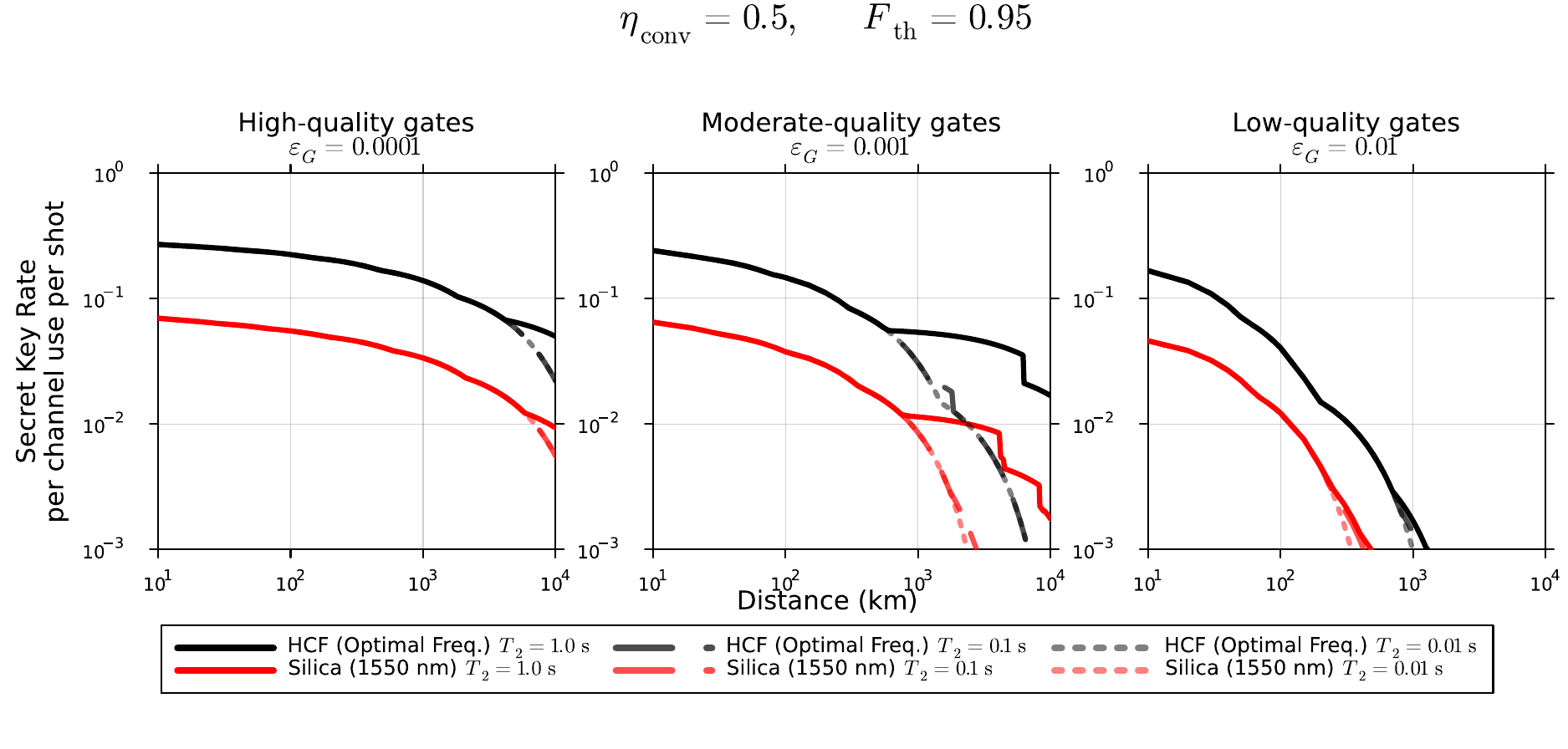}
    \caption{\textbf{Secret-key rate per channel use per shot (SKR/PCU) as a function of total distance for HCF and silica SMF with $\eta_\mathrm{conv}=0.5$.}
    Panels correspond to gate error probabilities (a) $\epsilon_G=10^{-4}$, (b) $\epsilon_G=10^{-3}$, and (c) $\epsilon_G=10^{-2}$. Black curves denote HCF operated at the optimal wavelength (the better of the memory-native $\sim780\,\mathrm{nm}$ and telecom 1550\,nm), while red curves denote silica SMF at 1550\,nm. Line styles indicate memory coherence times $T_2\in\{1,0.1,0.01\}\,\mathrm{s}$. Across all regimes, HCF sustains higher key rates over longer distances, with the advantage becoming more pronounced as gate errors increase and memory coherence decreases.}
    \label{fig:skr_adaptive_t2sweep}
\end{figure}

We evaluate the performance of HCF and silica SMF repeater architectures using the secret-key rate per channel use (SKR). Figure~\ref{fig:perf_eval} reports the achievable SKR as a function of the total end-to-end distance between Alice and Bob under different hardware regimes. Columns correspond to gate-quality regimes ($\epsilon_G = 10^{-4}, 10^{-3}, 10^{-2}$), while rows correspond to two-stage aggregate frequency-conversion efficiencies of $\eta_\mathrm{conv}=1.0$ and $\eta_\mathrm{conv}=0.5$. In each panel we compare HCF operating at its optimal wavelength with silica transmission at 1550\,nm, as well as HCF constrained to telecom operation.

Across all regimes shown in Fig.~\ref{fig:perf_eval}, HCF-based architectures achieve higher secret-key rates than silica-based designs. The advantage becomes increasingly pronounced at long distances and when frequency-conversion efficiency is limited. Even when HCF is constrained to operate at $1550\,\mathrm{nm}$ (gray dashed curves), its performance remains comparable to or better than silica while permitting similar or larger repeater spacing for equivalent SKR targets. When HCF operates at its optimal wavelength (black curves), both the achievable SKR and the feasible communication distance improve further. This advantage arises from the combined effects of favorable attenuation characteristics, reduced propagation delay, and the ability to exploit memory-native transmission wavelengths when advantageous.

To further examine whether the observed advantage depends on the additional frequency-conversion stage required for short-wavelength silicon APD-based detection, we compare against an alternative architecture using direct telecom-band SNSPDs (Fig.~\ref{fig:snsdp_comparison}). In this configuration, photons are converted once from the memory wavelength to $1550\,\mathrm{nm}$ and detected directly using SNSPDs, eliminating the second frequency-conversion stage. Direct telecom detection improves the performance of both HCF and silica SMF by reducing conversion losses. However, even when silica is equipped with this telecom detection architecture, HCF operated with the original APD-based architecture continues to outperform silica across the considered hardware regimes. In an optimized HCF implementation, where detector technology is also selected as an architectural parameter, direct telecom SNSPD detection can further improve the achievable performance, particularly in regimes with lower-quality local operations. These results confirm that the HCF advantage is not an artifact of the additional frequency-conversion stage required for telecom transmission with APD detection. Instead, it reflects the combined benefits of favorable HCF attenuation characteristics, reduced propagation delay, and the ability to exploit memory-native transmission when advantageous. These effects become increasingly important at longer distances, where entanglement distribution is limited by both transmission loss and memory decoherence.

These trends are summarized more systematically in Fig.~\ref{fig:skr_ratio_conv_eff}, which plots the ratio $\mathrm{SKR}_{\mathrm{HCF,optimal~freq.}}/\mathrm{SKR}_{\mathrm{silica}}$ as a function of total distance and conversion efficiency. The ratio exceeds unity across nearly the entire parameter space for the gate-error regimes considered, indicating that adaptive HCF consistently outperforms silica-based transmission. Only a small region at short distances and near-unity conversion efficiency approaches parity between the two architectures. Figure~\ref{fig:skr_adaptive_t2sweep} further examines the impact of memory coherence time. The advantage of HCF persists across all gate-quality regimes even for shorter memory lifetimes, demonstrating that the transmission benefits of HCF translate robustly into architectural improvements for quantum repeater networks. Taken together, these results show that HCF enables higher achievable key rates and longer repeater spacing, particularly in regimes where conversion losses or hardware noise limit conventional fibers.

\subsection{Effect of hardware efficiency}\label{sec:results_hardwareeff}
\begin{figure}[!t]
\centering
    \begin{subfigure}{0.4\linewidth}
            \includegraphics[width=\linewidth]{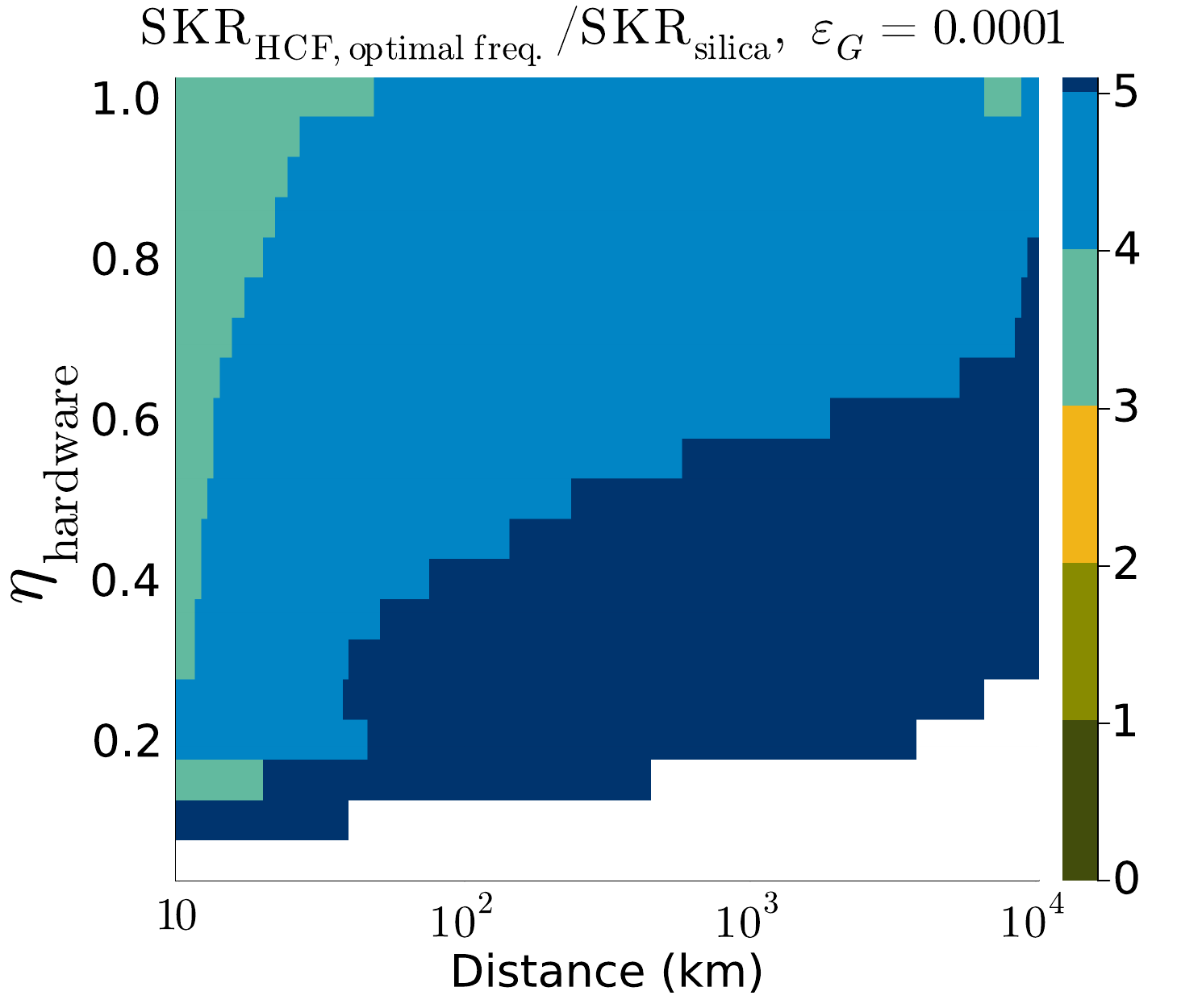}
    \end{subfigure}
        \begin{subfigure}{0.4\linewidth}
            \includegraphics[width=\linewidth]{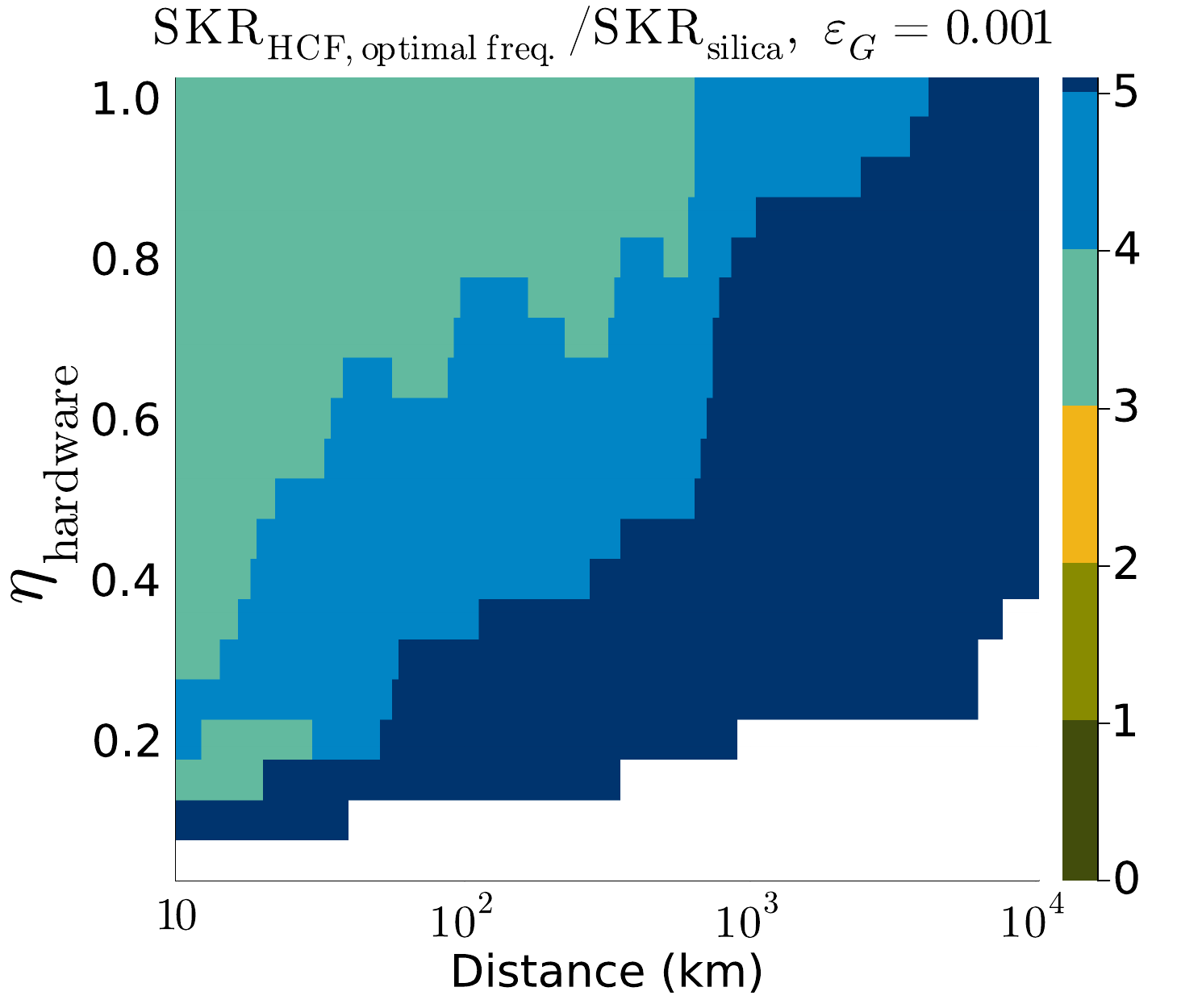}
    \end{subfigure}
    \caption{\textbf{Ratio $\mathrm{\textbf{SKR}}_{\mathrm{\textbf{HCF,opt}}}/\mathrm{\textbf{SKR}}_{\mathrm{\textbf{silica}}}$ as a function of total distance and hardware efficiency $\eta_{\mathrm{\textbf{hardware}}}$, with $\boldsymbol{\eta}_{\mathbf{conv}}=\mathbf{0.5}$.}
    The left and right panels correspond to $\epsilon_G=10^{-4}$ and $\epsilon_G=10^{-3}$, respectively. Colors denote binned values of the ratio, with larger values indicating stronger advantage for adaptive HCF. Across the plotted range, the ratio remains above unity, indicating that HCF matches or exceeds silica performance throughout. The advantage increases at longer distances and lower hardware efficiencies, where reduced attenuation allows HCF to sustain key generation while silica performance rapidly degrades. Regions where silica yields negligible key rate while HCF remains operational appear in the highest bin, while regions where both yield zero key rate are shown in white.}
    \label{fig:skr_ratio_hardwareff}
\end{figure}

In addition to fiber attenuation and frequency conversion losses, elementary link generation is affected by several hardware-related inefficiencies, including photon emission probability, detector efficiency, and coupling losses at the detection interface. In our model these factors are captured through the aggregate parameter $\eta_{\mathrm{hardware}}$, defined in Sec.~\ref{sec:system_description}. Physically, $\eta_{\mathrm{hardware}}$ represents the combined efficiency of photon emission, detection, and associated coupling losses that are independent of the transmission fiber technology.

To understand how these device-level inefficiencies influence architectural performance, we sweep $\eta_{\mathrm{hardware}}$ across a broad range of values while keeping the two-stage aggregate frequency conversion efficiency fixed at $\eta_\mathrm{conv}=0.5$. Figure~\ref{fig:skr_ratio_hardwareff} reports the ratio $\mathrm{SKR}_{\mathrm{HCF,optimal~freq.}}/\mathrm{SKR}_{\mathrm{silica}}$ as a function of total distance and hardware efficiency for two representative gate-error regimes.

Across the explored parameter space, adaptive HCF consistently matches or exceeds the performance of silica transmission. The advantage becomes increasingly pronounced at long distances and at lower hardware efficiencies, where the attenuation advantage of HCF allows entanglement generation to remain viable even as silica-based links rapidly lose performance. In regions where silica yields negligible key rate while HCF continues to generate entanglement, the ratio saturates in the highest bin of the heatmap. These results demonstrate that the architectural benefits of HCF persist even when realistic hardware inefficiencies are taken into account.

This sweep also bounds the sensitivity of our conclusions to the coupling model of Sec.~\ref{sec:memory_tofiber_coupling}. In Eq.~\eqref{eq:p_succ_elem}, $\eta_{\mathrm{hardware}}$ and $\eta_{\mathrm{coupling, memory\to fiber}}$ enter only through their product, so Fig.~\ref{fig:skr_ratio_hardwareff} is equivalently a sweep over total coupling efficiency. Note that it does not vary the difference in coupling between SMF and HCF, which enters as a fixed factor $(0.79/0.83)^2\approx0.91$ on the elementary-link success probability of HCF relative to SMF. Since the interface penalty is distance-independent while the transmission advantage grows with inter-repeater spacing $L_0$, no fixed asymmetry alters the conclusions at the spacings considered here.

\subsection{Operational costs}\label{sec:results_costs}
\begin{figure}[!t]
\centering
    \begin{subfigure}{0.45\linewidth}
            \includegraphics[width=\linewidth]{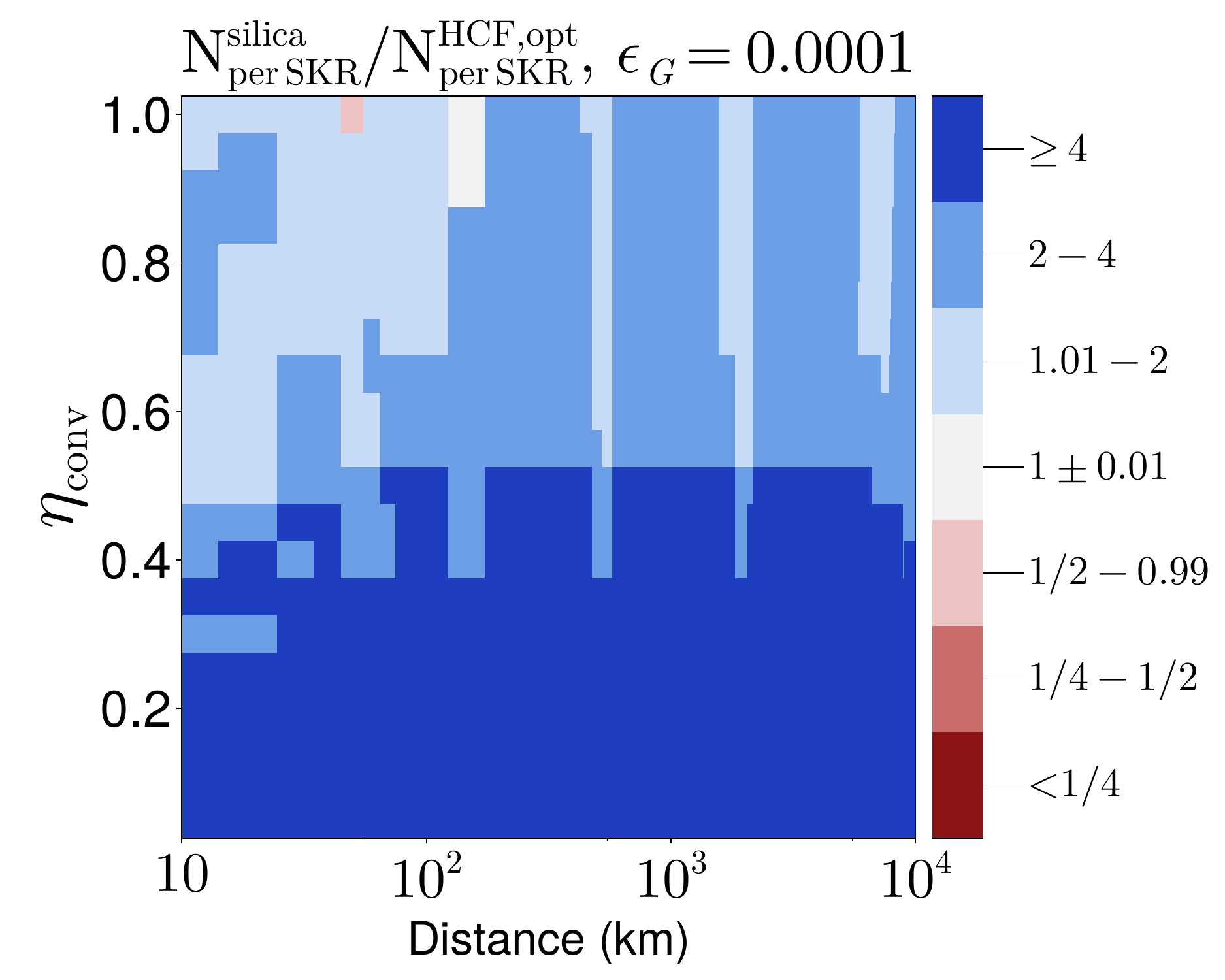}
    \end{subfigure}
    \begin{subfigure}{0.45\linewidth}
            \includegraphics[width=\linewidth]{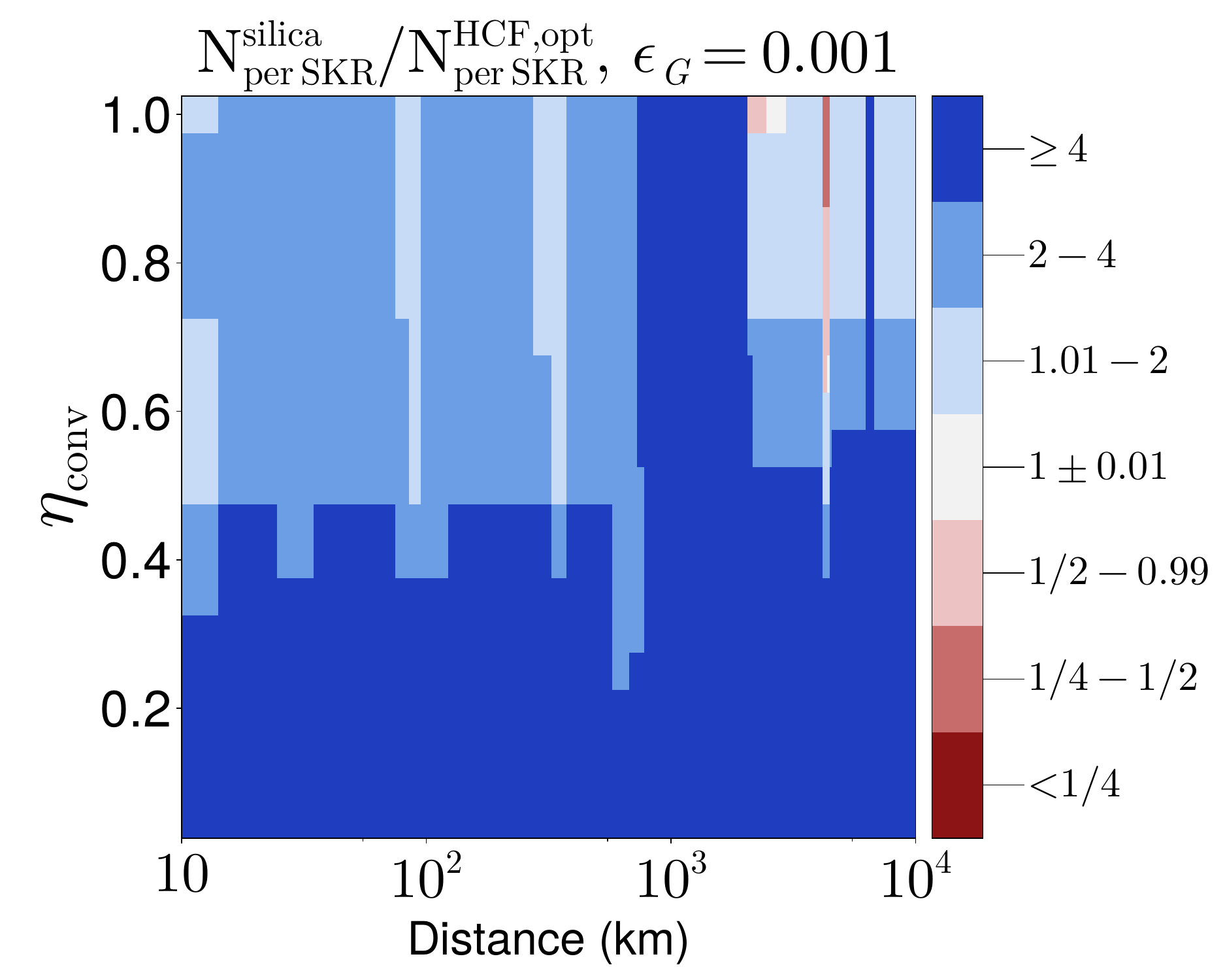}
    \end{subfigure}
    \caption{\textbf{Ratio of repeater nodes required per unit secret-key rate, $N_{\mathrm{per\,SKR}}^{\mathrm{silica}}/ N_{\mathrm{per\,SKR}}^{\mathrm{HCF,opt}}$, as a function of total distance and conversion efficiency.} The left and right panels correspond to $\epsilon_G=10^{-4}$ and $\epsilon_G=10^{-3}$, respectively. Colors denote binned values of the ratio: blue ($>1$) indicates that silica requires more repeater nodes per unit secret-key rate (favoring HCF), red ($<1$) indicates the opposite, and the central band denotes near parity. Across most of the parameter space, the ratio exceeds unity, indicating that HCF achieves comparable secret-key throughput with fewer repeater nodes. The advantage becomes more pronounced at longer distances and moderate conversion efficiencies, where reduced loss allows HCF to maintain key generation with larger repeater spacing.}
    \label{fig:N_ratio_conv_eff}
\end{figure}

\begin{figure}[!htb]
\centering
    \centering
    \begin{subfigure}{0.95\linewidth}
    \caption{Perfect Conversion Efficiency}
    \includegraphics[width=\linewidth,
                         trim=0cm 2cm 0cm 2cm, clip]{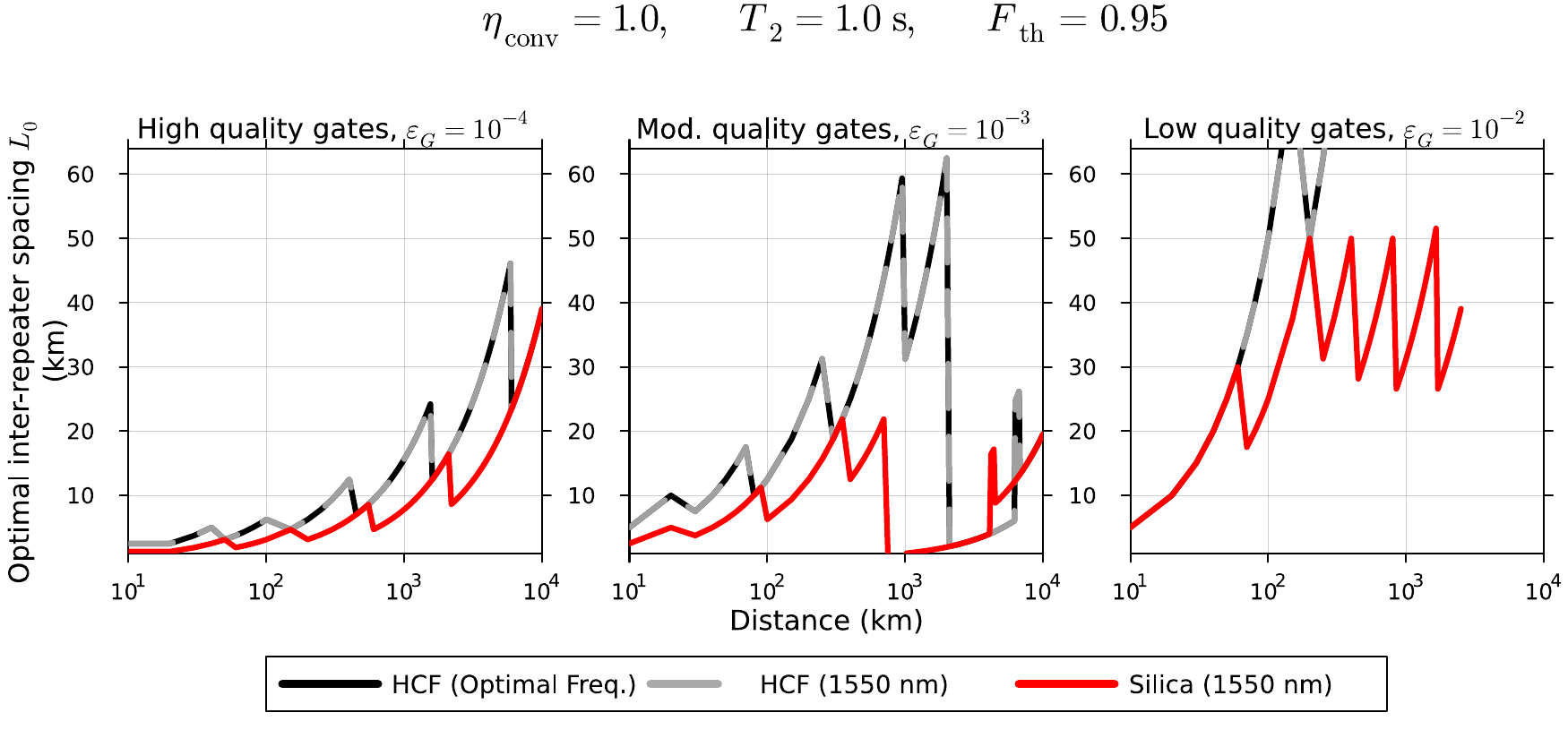}
\end{subfigure}
    \begin{subfigure}{0.95\linewidth}
    \caption{Moderate Conversion Efficiency}
    \includegraphics[width=\linewidth,
                         trim=0cm 0cm 0cm 2cm, clip]{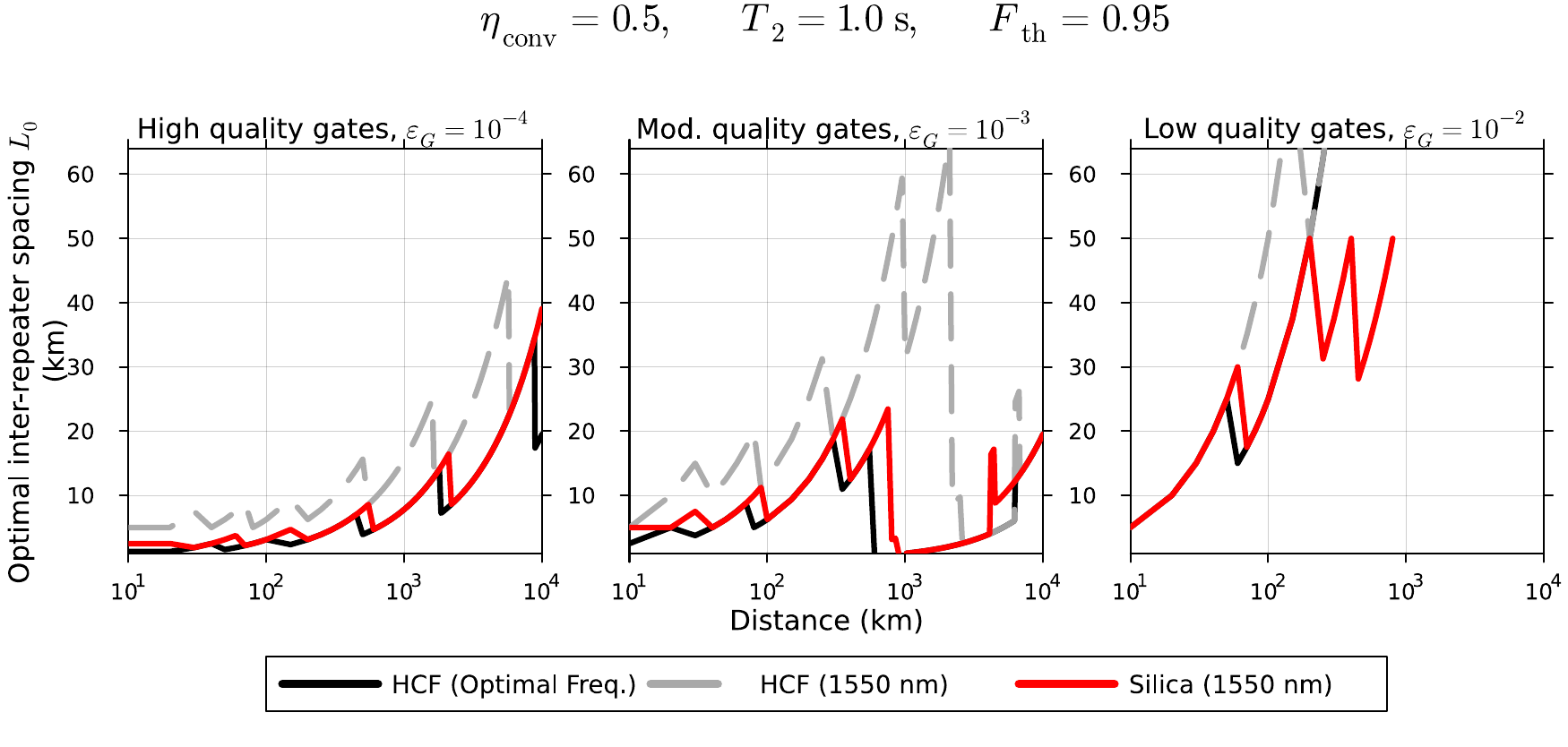}
    \end{subfigure}
    \caption{
    \textbf{Optimal inter-repeater spacing $L_0$ required to maximize the secret-key rate for HCF and silica SMF quantum repeater architectures as a function of the total end-to-end distance between Alice and Bob.} Columns correspond to gate-quality regimes (high quality -- $\epsilon_G = 10^{-4}$, medium quality -- $\epsilon_G = 10^{-3}$, and low quality -- $\epsilon_G = 10^{-2}$), illustrating how the optimal repeater spacing changes as local operations become noisier. Rows correspond to the assumed two-stage aggregate frequency conversion efficiency: perfect conversion ($\eta_\mathrm{conv}=1.0$, top row) and moderate conversion ($\eta_\mathrm{conv}=0.5$, bottom row).
    Black curves denote HCF operating at the \emph{optimal wavelength}, defined as the choice between the memory-native emission wavelength ($\sim$780\,nm) and telecom wavelength (1550\,nm) that maximizes the elementary entanglement generation probability for the link. Gray dashed curves denote HCF operation constrained to telecom wavelength (1550\,nm), while red curves denote conventional silica SMF operation at 1550\,nm.
    Across all gate-quality regimes, HCF generally supports larger optimal repeater spacings than silica SMF, particularly at long distances and when frequency conversion efficiency is limited. This indicates that HCF-based links can achieve comparable end-to-end performance with fewer repeaters. Even when restricted to telecom operation at 1550\,nm, HCF frequently supports equal or larger repeater spacing than SMF while delivering comparable or improved performance. }\label{fig:interrepeater}
\end{figure}

Since the repeater architectures considered here employ the same underlying hardware components, the base repeater design and associated hardware costs are largely comparable for HCF and silica implementations. Instead, differences in operational cost arise primarily from the number of repeater nodes required to sustain a given level of secret-key throughput.
{To quantify this effect, we define
\begin{align*}
    N{_\mathrm{per\,SKR}}=\frac{N}{\mathrm{SKR}},
\end{align*}
where $N$ is the total number of repeater nodes required for the end-to-end connection. This metric captures the repeater infrastructure and associated operational overhead required per unit secret-key rate.

The resulting repeater-resource comparison is summarized in Fig.~\ref{fig:N_ratio_conv_eff}, which plots the ratio of repeater nodes required per delivered secret key between silica SMF and HCF architectures. Across most of the explored parameter space this ratio exceeds unity, indicating that SMF-based networks require more repeater nodes to achieve the same secret-key throughput. The advantage of HCF becomes particularly pronounced at longer distances and moderate conversion efficiencies, where improved transmission performance allows HCF links to maintain key generation with larger inter-repeater spacing.

Beyond repeater count, other resource metrics exhibit similar trends, including the number of required qubits, two-qubit gate operations, and measurement operations (see Supplementary Materials~\cite{mantri-prateekMantriprateekRethinkingQuantumNetworkingwithAdvancesinFiberTechnology2026}).

Figure~\ref{fig:interrepeater} shows the corresponding optimal inter-repeater spacing that maximizes SKR for both fiber technologies across different hardware regimes. Consistent with the trends above, HCF generally supports larger inter-repeater spacing than silica, particularly at long distances and when conversion efficiency is limited, which directly translates into fewer repeaters required for a given end-to-end link.

Together, these results indicate that the transmission advantages of HCF translate not only into higher achievable key rates but also into lower operational resource requirements for long-distance quantum repeater networks.


\section{Conclusion and Outlook}\label{sec:Conclusions}

We evaluated the impact of recent hollow-core fiber (HCF) advances on multiplexed two-way quantum repeater networks and compared their performance against conventional silica single-mode fiber (SMF) links across a range of operating regimes. Our results show that HCF provides a consistent architectural advantage over SMF when end-to-end performance is measured through secret-key rate per channel use, feasible repeater spacing, and operational resource costs.

In particular, HCF operating at the optimal transmission wavelength achieves substantially higher secret-key rates than silica-based transmission, with the advantage becoming especially pronounced at long distances and under limited frequency-conversion efficiency. Even when constrained to telecom-band operation, HCF remains competitive with or superior to SMF while frequently enabling larger optimal inter-repeater spacing. This translates directly into sparser repeater deployment and lower operational overhead for a given end-to-end secret-key throughput. We further showed that these benefits persist under realistic noise, including finite memory coherence time, imperfect hardware efficiency, and imperfect two-qubit gate operations. Taken together, these results indicate that recent advances in HCF materially expand the design space of terrestrial quantum repeater networks.

An implication of this study is that the transmission medium should not be treated as a fixed background assumption in terrestrial quantum-network design. In conventional silica-based architectures, operation is strongly biased toward telecom wavelengths because of the favorable attenuation window of SMF. However, many leading quantum memory platforms and several attractive detector technologies do not naturally operate in this band. As a result, silica-based systems often incur additional penalties through frequency conversion, insertion loss, and interface complexity. By reducing propagation loss and broadening the range of viable transmission wavelengths, HCF opens the possibility of architectures that are better aligned with memory-native operation. In that sense, HCF is not simply a lower-loss replacement for silica fiber; it changes the systems-level tradeoff among wavelength choice, conversion overhead, repeater spacing, and end-to-end performance.

These advantages should also be interpreted in light of cost. At present, HCF is less mature than conventional silica fiber and carries higher fabrication, integration, and deployment costs in the near term. Additional engineering challenges may exist in splicing, packaging, and stable interfacing. From the perspective of classical optical communications, such considerations are often decisive because the transmission medium itself constitutes a major fraction of the deployed infrastructure.

Quantum repeater networks, however, operate in a different cost regime. In these systems, the dominant costs are unlikely to arise from fiber alone, but from the repeater hardware deployed along the route: quantum memories, photonic interfaces, frequency-conversion modules, detectors, cryogenic systems where needed, control electronics, and the repeater stations themselves. Reducing the number of repeaters, enlarging the feasible repeater spacing, and lowering the operational workload required per delivered secret key may be more consequential than modest differences in the cost of the transmission fiber. Our results suggest that even if HCF remains more expensive than silica at the component level, it may still be the more attractive architectural choice if it sufficiently reduces repeater count or improves performance enough to lower the total system cost.

This point is particularly important because repeater spacing affects more than the number of devices in the chain. Fewer repeater stations also reduce the demands associated with deployment, calibration, synchronization, site access, and long-term maintenance. Since each repeater may require expensive quantum hardware and substantial operational support, the ability of HCF to sustain larger optimal inter-repeater spacing is not merely a transmission benefit; it may also improve the practical and economic feasibility of terrestrial quantum communication.

Our model also neglects Raman scattering, the dominant limit on coexistence of classical and quantum channels in solid-core SMF. Since the Raman contribution scales with the overlap between the guided mode and silica, antiresonant HCF suppresses it substantially. DV-QKD has been demonstrated coexisting with 1.6~Tbps of classical traffic at 0~dBm total coexistence power in NANF, a power 250 times greater than that at which SMF already loses 73\% of its secret key rate~\cite{alia21cqhcf}. More recently, entanglement distribution has been established over 11.5~km of NANF alongside 800~Gbps of DWDM classical traffic spaced 1.2~nm away in the same band, with secret key rates preserved across a four-node network for 55 hours~\cite{clark25coexhcf}, reflecting the reduced thermal and acoustic sensitivity of an air-guided mode. Since our model omits these effects, the HCF advantages reported here would only grow in deployments where quantum and classical traffic coexist in the same fiber.

The HCF attenuation values used here are projected rather than measured. The gap is modest at 1550\,nm, where the measured $0.091$\,dB/km still gives $1.65\times$ the SMF attenuation length and leaves our telecom-band conclusions intact. Similarly, our memory-native results anticipate continued progress in that band to achieve the predicted value of $0.180$~dB/km, following a trajectory that has already brought 1550\,nm attenuation from $0.174$ to $0.091$\,dB/km within three years \cite{Petrovich2025}.

At the same time, several directions remain open for future work. A natural next step is a fuller techno-economic analysis that explicitly compares fiber deployment costs with repeater-node costs, including memories, detectors, conversion hardware, and associated control systems. Such a study would help identify the parameter regimes in which HCF is preferable on a total-cost basis rather than purely on a performance basis. It would also be valuable to extend the present analysis to explicit models of classical--quantum coexistence, where the reduced nonlinear noise of HCF may provide additional system-level advantages. More detailed experimental characterization of HCF coupling, packaging, and photonic interface losses will be important for determining how much of the modeled benefit can be realized in practical systems. 
The fiber-managed interface assumed in Sec.~\ref{sec:memory_tofiber_coupling} could be simplified considerably by coupling directly from chip to hollow core which has yet to be demonstrated and would require edge-coupler mode-field diameters well beyond current designs. Further, optimal inter-repeater spacings reported here exceed current reported draw lengths of $10$-$15$~km, and would therefore require splicing of HCF segments which is less mature than SMF fiber splicing \cite{Zhang26splicing}.

Overall, the present study shows that recent progress in hollow-core fiber can do more than improve a single physical link: it can shift the preferred operating regime of practical terrestrial quantum repeater networks. For quantum networks, where repeaters, memories, and associated hardware are likely to dominate total cost, the ability of HCF to improve rates while reducing repeater requirements makes it a particularly promising direction for future experimental and commercial development. 

More broadly, progress in quantum-dot single-photon sources~\cite{senellart2017high, carosini2026quantum}, low-loss electro-optic programmable interferometers enabling ultrafast switching~\cite{sund2023high, dong2023synchronous}, and efficient near-infrared single-photon detection~\cite{an2025,excelitas} highlights the need to design quantum networks around the wavelengths and interfaces favored by quantum hardware. With the advances in HCF, these developments motivate treating the fiber platform and operating wavelength as co-optimized architectural degrees of freedom rather than inheriting the 1550~nm operating point of classical communications.

\section*{Competing Interests}\label{sec:competing_interests}
The authors declare no competing interests.

\section*{Acknowledgments}\label{sec:acknowledgment}
We thank the anonymous reviewers for their careful reading and constructive comments, which helped improve the manuscript. We also thank the Manning College of Information and Computer Sciences at the University of Massachusetts Amherst for providing access to its High Performance Computation Cluster. D.T., P.M., and M.S.B. acknowledge funding support from the NSF- ERC Center for Quantum Networks grant EEC-1941583, NSF grant CNS-2402861,  and DOE Grant AK0000000018297. R.N. acknowledges the support of the Riccio College of Engineering, University of Massachusetts Amherst. R.N. and D.T. gratefully acknowledge support from the Office of the Provost and Paros Center for Atmospheric Research, University of Massachusetts Amherst. 
\appendix
\renewcommand{\thesection}{\Alph{section}}
\renewcommand{\thesubsection}{\thesection.\Roman{subsection}}
\renewcommand{\theequation}{\thesection.\arabic{equation}}
\begin{appendices}

\section{Recursive formulation of the probability distribution}\label{appendix:recursive_calcs}

The material for this section is taken from Mantri \textit{et al.}~\cite{mantri2025comparing}. We consider a linear network with $N=2^n$ links. Let $M = m 2^{n + 1}$ denote the number of multiplexed channels available at each elementary link in a single shot, with $m \ge 1$. This assumption can be relaxed when distillation is not performed at every level, in which case $M$ may be taken to be lesser than $m 2^{n + 1}$. We consider a nested pumping distillation protocol that performs at most one distillation round at each level.

Let $Y_i$ denote the number of Bell pairs on a segment at level $i$ with $p_{i,k} = P(Y_i = k)$ denoting the probability of having exactly $k$ Bell pairs at level $i$. Let $\pi_0$ denote the probability that a link-level Bell pair generation attempt succeeds, and let $\overline{\pi}_0 = 1 - \pi_0$. Then,
\begin{align}\label{eq:basic_prob_generation}
    p_{0,k} = \binom{M}{k} \pi_0^k \overline{\pi}_0^{M - k}, \quad k= 0,1,\ldots, M\ .
\end{align} 

\noindent The effect of a distillation operation at level $i$ whenever it is performed is captured by $q_{i,k}$,
\begin{align}
    q_{i,k} = P(X_i = k) = \sum_{j=2k}^{M/2^i} p_{i,j} \binom{\lfloor j/2 \rfloor}{k} d_i^k \overline{d}_i^{\lfloor j/2 \rfloor - k}, \quad i \in \{0,\cdots, n\} ;k = 0,1,\ldots, \lfloor M/2^{i+1} \rfloor.
\end{align}
where $X_i$ denotes the number of Bell pairs produced by one distillation step at level $i$ when performed, and $d_i$ the probability of a successful distillation step. In the absence of distillation at  level $i$, $ q_{i,k}=p_{i,k}$. Hence
\begin{align}
    q_{i,k} = P(X_i = k) = 
    \begin{cases}
            \sum_{j=2k}^{M_i} p_{i,j}\binom{\lfloor j/2 \rfloor}{k} d_i^k \overline{d}_i^{\lfloor j/2 \rfloor - k}, &\mathcal{D}_i = 1 \\ 
            p_{i,k}, & \mathcal{D}_i = 0 
    \end{cases},\quad i=0,\ldots,n ;k  = 0,1, \ldots,\lfloor M_i/2 \rfloor
\end{align} 
where $M_i = \lfloor M / 2^{\sum_{j=0}^{i-1}\mathcal{D}_j} \rfloor$ for $i > 0$, and $\mathcal{D}_i = 0$, if no distillation is performed at level $i$, and $\mathcal{D}_i = 1$ otherwise. As considered in Mantri\textit{ et al.}~\cite{mantri2025comparing}, we set $\mathcal{D}_n = 0$, i.e., no end-to-end distillation is allowed. Now,

\begin{align}
  p_{i,k} = (q_{i-1, k})^2 + 2q_{i-1, k} \sum_{j= k+1}^{M_{i}}q_{i-1,j}, \quad i=1,\ldots,n; k = 0,1, \ldots, M_i.
\end{align}
The protocol considered here terminates whenever $Y_i < 1$. 
Let,
\begin{align*}
    r_i &= \text{Probability of reset at level $i$ conditioned on reaching level $i$},\\
    p'_{i,k} &= P(Y_i = k | \text{ no reset at levels } 0,1,\dots,i),\\
    q'_{i,k} &= \text{Probability of having $k$ distilled pairs conditioned on reaching level $i$}.
\end{align*}
We initialize these values as,
\begin{align}
        r_0 &= \sum_{j=0}^{R_0-1} \binom{M}{j} \pi_0^j \overline{\pi}_0^{M - j} \quad, \text{where }  R_0 = 1\\\nonumber
        &= \overline{\pi}_0^{M }
\end{align}
and
\begin{align}\label{eq:p_prime_0}
    p'_{0,k} = 
    \begin{cases}
        \big(\binom{M}{k}\pi_0^k \overline{\pi}_0^{M - k}\big)/(1 - r_0), &  k \ge 1\\
        0, &   k < 1
    \end{cases}, \quad k \in \{0,1, \dots, M\}. 
\end{align}

Now,
\begin{align}\label{eq:q_prime_dejmps}
    q'_{i,k} &= P(X_i = k | \text{ no reset at levels } 0,1,\dots,i),\nonumber\\
    &= \begin{cases}
             \sum_{j = 2k}^{M_i} p'_{i,j} \binom{\lfloor j/2\rfloor}{k}d_i^k \overline{d}_i^{\lfloor j/2 \rfloor - k}, & \mathcal{D}_i = 1 \\
            p'_{i,k}, &\mathcal{D}_i = 0 \\
    \end{cases}, \quad k \in \{0,1,\dots, \lfloor M_i/2 \rfloor\}.
\end{align}

\begin{align}\label{eq:conditional_reset_prob}
    r_i = 
    \begin{cases}
        \sum_{l=0}^1\big((q'_{i-1, l})^2 + 2q'_{i-1, l} \sum_{j = l+1}^{M_{i}} q'_{i-1,j}\big), & \mathcal{D}_i = 1 \\
       0, & \mathcal{D}_i = 0 \\
    \end{cases}; i \in \{1, \dots, n\},
\end{align} 
and,
\[
\label{eq:p_prime_i}
    p'_{i,k} = 
        \begin{cases}
            \big((q'_{i-1, k})^2 + 2q'_{i-1, k} \sum_{j= k+1}^{M_{i}}q'_{i-1,j} \big)/(1 - r_i) , & k \ge 1  \\         
            0 , & k < 1
        \end{cases}, 
        \quad k = 0, \dots, M_i; i= 1,\dots , n.
\] 

\noindent Let $f_i$ be the probability of reset at level $i$, 
\begin{align}\label{eq:reset_probability}
f_i =
    \begin{cases}
        1 - (1 - r_0)^N, & i = 0 \\
        (1 - (1 -r_i)^{{\frac{N}{2^i}}})\prod_{j = 0}^{i-1} (1 - r_j)^{\frac{N}{2^j}}, &\text{otherwise}\\
    \end{cases}
\end{align}\\
In this manuscript, we compute the decision to distill $\mathcal{D}_i$ using a fidelity threshold metric $ F_{th}$, with distillation being performed when the condition $F_i < F_{th}$ holds, where $F_i$ is the local fidelity of the state at nesting level $i$. For our analysis we consider a fidelity threshold $F_{th} = 0.95$.

\end{appendices}

\printbibliography
\end{document}